\documentclass{styling/nseJournal}

\usepackage{styling/defs}

\begin{document}

\title{Multigroup Neutron Transport using a Collision-Based Hybrid Method} 

\addAuthor{Ben Whewell}{a}
\addAuthor{\correspondingAuthor{Ryan G.~McClarren}}{a}
\correspondingEmail{rmcclarr@nd.edu}
\addAuthor{Cory D. Hauck}{b}
\addAuthor{Minwoo Shin}{a}

\addAffiliation{a}{Department of Aerospace and Mechanical Engineering \\ University of Notre Dame, Notre Dame, IN 46556, USA}
\addAffiliation{b}{Computer Science and Mathematics Division \\ Oak Ridge National Laboratory, Oak Ridge, TN 37830, USA}

\addKeyword{Neutron Transport}
\addKeyword{Hybrid Methods}
\addKeyword{Multigroup Approximation}

\titlePage

\begin{abstract}
A collision-based hybrid algorithm for the discrete ordinates approximation of the neutron transport equation is extended to the isotropic multigroup setting. 
The algorithm uses discrete energy and angle grids at two different resolutions and approximates the fission and scattering sources on the coarser grids.
The coupling of a collided transport equation, discretized on the coarse grid, with an uncollided transport equation, discretized on the fine grid, yields an algorithm that, in most cases, is more efficient than the traditional multigroup approach.
The improvement over existing techniques is demonstrated for time-dependent problems with different materials, geometries, and energy groups.
\end{abstract}

\section{Introduction}
The neutron transport equation (NTE) is used to model neutron populations traveling through different media.
A common computational technique used to solve this equation numerically is the discrete ordinates (S$_N$) method~\cite{lewis1993, chandrasekhar2013}.
In a multigroup discrete ordinates calculation, time-dependent problems are often discretized implicitly.  This results in a steady-state problem at each time step in the computation that is solved iteratively.  Typically  there are two levels of iteration.
The inner iteration updates the scalar flux inside each energy group using a fixed-point source iteration scheme, where the in-group scattering source is lagged and each update requires the inversion of the operator that models streaming and loss. 
These updates, called transport sweeps, amount to the inversion of a block triangular system in space for each angle \cite{o1987transport}.   The outer iterations over the energy variable often follow a Gauss-Seidel strategy that updates the scattering and fission source terms as the energy groups are updated in order from highest to lowest energies~\cite{bell1970nuclear}.
The Gauss-Seidel method is effective in problems that are dominated by down scattering; however, it is dependent on the number of interactions that the neutrons have with the material, such as scattering and, in time-dependent problems, fission.
For example, a purely down-scattering material without fission will converge in one Gauss-Seidel iteration. If there is up-scattering or fission, the number of iterations can become prohibitively large in optically thick problems \cite{lewis1993}.

The cost of a multigroup calculation is a function of the energy resolution needed; coarsening the energy grid, thus reducing the number of energy groups, can lead to less expensive solutions.
However, a smaller energy group structure must be carefully chosen to limit the discretization error and preserve the characteristics of the energy-dependent cross section~\cite{njoy}. 
Likewise, there can be faster convergence times using a low number of angles, but this can also lead to both larger errors and exacerbate the non-physical ray effects that can arise in S$_N$ calculations. 
There are techniques for reducing these ray effects~\cite{hauck2019filtered, frank2020}, but the most common solution is to increase the number of angles~\cite{lewis1993}, thus increasing the computational time.

Acceleration techniques, such as coarse mesh rebalancing and diffusion synthetic acceleration, are used to reduce the computational cost of the iterative solver.
Unfortunately, these methods cannot be implemented indiscriminately, as coarse mesh rebalancing must be concerned with the coarse mesh size~\cite{lewis1993} while diffusion synthetic acceleration has difficultly in highly heterogeneous materials~\cite{southworth2021}. 
For the outer iterations, there are two-grid, nonlinear diffusion acceleration, and Krylov subspace schemes that can improve the convergence of problems with upscattering or fission \cite{adams1993two, anistratov2013multilevel, slaybaugh2018eigenvalue}. 
While these methods demonstrate improvements with high upscattering materials, there is a nominal improvement over Gauss-Seidel in low upscattering materials~\cite{adams1993two}. 
These methods must also include the coarse grid diffusion equation solver for their transport code bases, something that is not needed for the collision-based hybrid method.
Another approach would be to use a higher-order time integration method \cite{anistratov2013multilevel,edwards2011nonlinear,mcclarren2008quasi,thoreson2009high}. Nevertheless, the robustness of the fully-implicit backward Euler method makes it a standard choice for transport problems \cite[\S2.5]{lewis1993}.

More recently, collision-based hybrid algorithms for time-dependent transport equations have been provender for transport researchers \cite{hauck2013,crockatt2017arbitrary,crockatt2019hybrid,crockatt2020improvements,heningburg2020}.  These algorithms split the transport equation into collided and uncollided components, as was done much earlier in the steady-state context \cite{alcouffe1977}.
In the aforementioned examples, the hybrid approach yielded substantial improvements in efficiency when compared to monolithic discretization schemes.
While previous work has applied different spatial discretization schemes to the collided and uncollided components, the approach has so far only been implemented on mono-energetic problems.
In the current work, we extend the collision-based hybrid algorithm to the multigroup setting. Specifically, we apply the algorithm to the multigroup discrete ordinates approximation of the time-dependent neutron transport equation in one-dimensional problem geometries. 
We test the algorithm on one-dimensional slabs and spheres with different boundary conditions and  sources. We find that the hybrid algorithm can for the same computational time deliver much smaller error or reach the same error level produce solutions in significantly less time. Additionally, we find that the applying a collided-uncollided splitting to a monolithic discretization grid recovers the same solution as standard Gauss-Seidel in less time.  This improvement in time-to-solution is due to the changes in the solver that reduce unnecessary iterations in the inner loop of the nested approach.  A more focused study of oversolving with nested iterations and how to avoid it can be found in \cite{senecal2017}.

The remainder of this paper is organized as follows.  In Section~\ref{sec:neutron-transport}, we introduce the neutron transport equation, the continuous collided-uncollided split, the multigroup discrete ordinates transport equations, and the collision-based hybrid algorithm. In Section~\ref{sec:results}, we present numerical results for three different test problems and compare the hybrid method to more traditional approaches. In Section~\ref{sec:conclusion}, we present conclusions and discuss future work.

\section{The neutron transport equation and the hybrid formulation} \label{sec:neutron-transport}
The NTE models the behavior of advecting neutrons that interact with a surrounding material medium.  It does so by tracking the evolution of the angular flux $\Psi$ which depends on a spatial coordinate $\bx \in D \subset \bbR^3$, angular coordinate $\bsOmega \in \bbS^2$, energy $E>0$, and time $t>0$.  The NTE takes the form \cite{lewis1993} 
\begin{equation} \label{eq:transport-continuous} \begin{split}
    \left( \frac{1}{c(E)}\frac{\partial}{\partial t} + \Omega \cdot \nabla + \sig{t}(\bx, E) \right) \Psi(\bx,\bsOmega,E,t) &= 
    \int_{0}^{\infty}dE' \; \sig{s} (\bx, E' \rightarrow E,t)\overline \Psi(\bx, E', t)\\
    & \hspace{-30pt}+ \chi(\bx, E) \int_{0}^{\infty} dE' \; \nu(x, E') \sig{f}(\bx, E')  \overline \Psi(\bx, E', t) +  Q(\bx,\bsOmega, E, t),
\end{split} \end{equation}
where the scalar flux $\overline \Psi: \left(D\times\bbR^+\times\bbR^+\right) \to \bbR$ is given by \begin{equation} \label{eq:unit-sphere}
  \overline \Psi(\bx,E,t) = \frac{1}{4 \pi} \int_{\mathbb{S}^2} \Psi (\bx,\bsOmega,E,t) d \Omega,
\end{equation}
and $c(E)$ is the velocity of a neutron with energy $E$.
The three terms on the left hand side of Eq.~\eqref{eq:transport-continuous} are, in order, the time derivative term, streaming term, and removal term; the right hand side is comprised of the scattering, fission, and external sources.  The material is characterised by cross-sections $\sig{s}$, $\sig{f}$, and $\sig{t}$.  More precisely, $\sig{s}(\bx, E' \rightarrow E)$ is the differential scattering cross section for neutrons scattering from energy $E'$ to $E$;  $\sig{f} (\bx, E')$ is the fission cross section at energy $E'$; and $\sig{t}(\bx, E)$ is the total cross section at energy $E$.  Given a fission event, $\nu(\bx, E')$ is the average number of neutrons created from fission caused by a neutron at energy $E'$ while $\chi (\bx, E)$ is the probability density that a neutron produced from fission will have energy $E$.

The initial condition for Eq.~\eqref{eq:transport-continuous} is \begin{equation} \label{eq:continuous-initial}
    \Psi(\bx, \bsOmega, E, t=0) = f( \bx, \bsOmega, E) \qquad \text{for} \quad \bx \in D, \quad \bsOmega \in \bbS^2, \quad E>0,
\end{equation} 
where $f$ is given.  The incoming boundary data is prescribed as \begin{equation} \label{eq:continuous-bcs}
    \Psi(\bx, \bsOmega, E, t) = b( \bx, \bsOmega, E, t) \qquad \text{for} \quad \bx \in \partial D, \quad  \bn(\bx) \cdot \bsOmega < 0, \quad E>0, \quad t >0,
\end{equation} 
where $\bn(\bx)$ the unit outward normal at $\bx \in \partial D$, the boundary of $D$.

\subsection{The continuous collided-uncollided split}
The goal of the hybrid approach is to accelerate the computation of a numerical solution for Eq.~\eqref{eq:transport-continuous} relative to a monolithic discretization, while minimizing the subsequent loss in accuracy.  The formulation of the hybrid method is best understood at the continuous level as a splitting method.  Indeed, because it is linear,  Eq.~\eqref{eq:transport-continuous} can be separated into two equations where a distinct angular flux is calculated in each and the results added together to get the total angular flux.

The first equation governs the uncollided angular flux, $\Psiu$.  It takes the form
\begin{equation} \label{eq:transport-uncollided} \begin{split}
    \left( \frac{1}{c(E)}\frac{\partial}{\partial t} + \Omega \cdot \nabla + \sig{t}(\bx, E) \right) \Psiu(\bx, \bsOmega,E,t) &= \Qu(\bx,\bsOmega, E, t) ,
    \end{split} 
    \end{equation} 
with $\Qu = Q$.
The initial condition for $\Psiu$ is 
\begin{equation}
    \Psiu(\bx, \bsOmega, E, 0) = f( \bx, \bsOmega, E)  \qquad \text{for} \quad \bx \in D, \quad \bsOmega \in \bbS^2, \quad E>0,
\end{equation} 
and the incoming boundary data is \begin{equation}
    \Psiu(\bx, \bsOmega, E, t) = b( \bx, \bsOmega, E, t) \qquad
    \text{for} \quad \bx \in \partial D, \quad  \bn(\bx) \cdot \bsOmega < 0, \quad E>0, \quad t >0.
\end{equation}
The second equation governs the collided angular flux, $\Psic$.  It takes the form
\begin{equation} \begin{split}
    \left( \frac{1}{c(E)}\frac{\partial}{\partial t} + \Omega \cdot \nabla + \sig{t}(\bx, E) \right) \Psic(\bx,\bsOmega,E,t) 
    &=  \int_{0}^{\infty}dE'  \sig{s} (\bx, E' \rightarrow E,t) \barPsic(\bx, E', t) \\
    &  \hspace{-30pt} +\chi(\bx, E) \int_{0}^{\infty} dE'  \nu(\bx,E') \sig{f}(\bx, E')  \barPsic(\bx, E', t)
  + \Qc(\bx,E, t),
\end{split} \end{equation} 
where the isotropic term $\Qc$ comes from the scattering and fission source of the uncollided flux:
\begin{equation}
     \Qc(\bx, E, t) = \int_{0}^{\infty}dE' \; \sig{s} (\bx, E' \rightarrow E,t) \barPsiu(\bx, E', t) + \chi(\bx, E) \int_{0}^{\infty} dE' \; \nu(\bx, E') \sig{f}(\bx, E')  \barPsiu(\bx, E', t).
\end{equation}
The initial condition for $\Psic$ is 
\begin{equation}
    \Psic(\bx, \bsOmega, E, 0) = 0 \qquad \text{for} \quad \bx \in D, \quad \bsOmega \in \bbS^2, \quad E>0,
\end{equation} 
and the incoming boundary data is \begin{equation}
    \Psic(\bx, \bsOmega, E, t) = 0 \qquad \text{for} \quad \bx \in \partial D, \quad \bn(\bx) \cdot \bsOmega < 0, \quad E>0, \quad t >0.
\end{equation}
A third equation for the total angular flux $\Psit$ takes the form
\begin{equation} \begin{split}
\label{eq: total_angular_flux}
    \left( \frac{1}{c(E)}\frac{\partial}{\partial t} + \Omega \cdot \nabla + \sig{t} \right) \Psit(\bx,\bsOmega,E,t) &= \Qt(\bx,\bsOmega, E, t),
\end{split} \end{equation} 
where the total external source $\Qt$ is \begin{equation} \begin{split}
    \Qt(\bx,\bsOmega, E, t) &= \Qc(\bx,\bsOmega, E, t) \\
        &+ \int_{0}^{\infty}dE' \; \sig{s} (\bx, E' \rightarrow E,t) \barPsic(\bx, E', t) +
    \chi(\bx, E) \int_{0}^{\infty} dE' \; \nu \sig{f}(\bx, E') \barPsic(\bx, E', t).
\end{split} \end{equation}
The initial condition for $\Psit$ is 
\begin{equation}
    \Psit(\bx, \bsOmega, E, 0) = f( \bx, \bsOmega, E), \qquad \text{for} \quad \bx \in D, \quad \bsOmega \in \bbS^2, \quad E>0,
\end{equation} 
and the incoming boundary data is 
\begin{equation}
    \Psit(\bx, \bsOmega, E, t) = b( \bx, \bsOmega, E, t) \qquad \text{for} \quad \bx \in \partial D, \quad \bn(\bx) \cdot \bsOmega < 0, \quad E>0, \quad t >0. 
\end{equation}

In the continuum formulation above,  uniqueness of solutions to the neutron transport equations implies that $\Psit = \Psiu + \Psic = \Psi$, making an independent equation for $\Psit$ trivially redundant.  However, this is no longer the case when equations for $\Psit$, $\Psiu$, and $\Psic$ are discretized with different methods and/or grids.  In practice, the hybrid method seeks at each time step a numerical approximation for $\Psiu$ on a fine grid and a numerical approximation for $\Psic$ on a course grid.  In the original formulation \cite{hauck2013}, these two approximate solutions were, at the end of the time step, combined via a remapping procedure in which an approximation for $\Psic$ is reconstructed on the fine grid.  Unfortunately, the reconstruction can introduce artifacts, particularly for $S_N$ discretizations in multi-dimensional settings.  To address this issue, the idea of using the equation for $\Psit$ to remap onto the fine grid was introduced in \cite{crockatt2020improvements} since the source $\Qt$ depends only on integrated quantities that can be computed on both grids.  An alternative view is that the collided/uncollided split provides a cheap method for approximating the scattering and, in this work, fission sources that appear in the right-hand side of Eq.~\eqref{eq:transport-continuous}.

The choices of initial conditions, boundary conditions, and sources for $\Psiu$ and $\Psic$ are not unique.  The strategy above is to assign data to the uncollided equation since it is equipped with the finest discretization.  However this strategy may not always be the best choice, particularly in highly-collisional problems with strong boundary layers \cite{Densmore2006cla}.

\subsection{The multigroup, discrete ordinates equations}
As an illustrative demonstration of the hybrid approach, we consider in this paper discretizations based on multigroup, discrete ordinates (S$_N$) equations \cite{chandrasekhar2013}.  We begin with a discussion of the monolithic method.
For the angular discretization, let $\bsOmega_m$ and $w_m$, where $ m \in \cM:=\{1, \cdots, M \}$, be discrete angles and weights for a quadrature rule over the sphere: that is, for any integrable function $u$ defined point-wise everywhere on $\bbS^2$,
\begin{equation}
    \frac{1}{4 \pi} \int_{\bbS^2} d \bsOmega\, u(\bsOmega) \approx \sum_{m=1}^M w_m u(\bsOmega_m).
\end{equation}
For the energy discretization,  $E_{\rm{max}}$ is a practical, finite upper bound on the neutron energy spectrum and let $0=E_0 < E_1 < \dots < E_G = E_{\rm{max}}$ be a set of finite energies that form the endpoints of $G$ non-overlapping intervals (or groups)  of width $\Delta E_g = E_{g} - E_{g-1}$.  Then define the function 
\begin{equation}
\psi_{m,g}(x,t) \approx \int_{E_{g-1}}^{E_{g}} dE \, \Psi(\bx, \bsOmega_m, E, t),\qquad g\in \cG:=\{1 ,\dots,G\}
\end{equation} 
as the solution of the multigroup, discrete ordinate (S$_N$) equations:
\begin{equation} \label{eq:transport-sn}
    \frac{1}{c_g} \frac{\partial }{\partial t} \psi_{m, g} + \mathbf{{\Omega}}_m \cdot \nabla \psi_{m, g} + \sig[g]{t}\psi_{m, g} = \sum_{g'=1}^{G} \sig[g' \rightarrow g]{s} \barpsi_{g'} +  \chi_{g} \sum_{g'=1}^{G} \nu_{g'} \sig[g']{f} \barpsi_{g'} + {q}_{m,g}, 
\end{equation}
where 
\begin{equation}
    \barpsi_{g} = \sum_{m=1}^M w_m \psi_{m,g}
\end{equation}
and the notation $g' \to g$ denotes scattering from energy group $E_{g'}$ to $E_g$.   The quantities $\sig[g]{t}$, $\sig[g' \rightarrow g]{s}$, $\chi_{g} $ and $\nu_{g'}$, and $\sig[g']{f}$ are all approximate weighted averages of their continuum counterparts.  For example,
\begin{gather}
    \sig[g' \rightarrow g]{s}(x, t) \approx \frac{\displaystyle\int_{E_{g-1}}^{E_{g}}  \displaystyle\int_{E_{g'-1}}^{E_{g'}} dE dE' \, \sig{s} (\bx, E' \rightarrow E, t) \overline \Psi (\bx, E', t)}{\displaystyle\int_{E_{g'-1}}^{E_{g'}} dE' \, \overline \Psi(\bx, E', t)} 
    \quand
         \chi_{g}(x) \approx \frac{\displaystyle\int_{E_{g-1}}^{E_{g}} dE \,  \chi(\bx, E) \overline \Psi (\bx, E, t)}{\displaystyle\int_{E_{g-1}}^{E_{g}} dE \, \overline \Psi(\bx, E, t)}.
\end{gather}
The approximation comes from the fact that $\overline \Psi(\bx, E, t)$ is not known a priori and an assumed spectral (and angular) shape of the solution must be used. 
In practice, these quantities are pre-calculated by nuclear data processing software such as NJOY \cite{njoy} or Fudge \cite{fudge} and, henceforth, are assumed to be given.
In our work the neutron velocity $c_g^{-1}$ is calculated from the relativistic energy formula from \cite[Figure 3]{bertozzi1964speed} at the midpoint of the energy bin. 

The initial condition for $ \psi_{m, g}$ is 
\begin{equation}
    \psi_{m, g} (\bx,0) = f_{m, g} (\bx) \qquad \text{for} \quad \bx \in D, \quad m \in \cM, \quad g\in\cG.
\end{equation} 
and the incoming boundary data is
\begin{equation}
    \psi_{m, g}(\bx,t) = b_{m, g} ( \bx, t)  \qquad \text{for} \quad \bn(\bx) \cdot \bsOmega_m < 0, \quad m \in \cM, \quad g\in\cG, \quad t > 0,
\end{equation} 
where 
\begin{equation}
    f_{m, g}(\bx) = \int_{E_{g-1}}^{E_g} dE \, f(\bx,\bsOmega_m,E)
    \qquand
     b_{m, g} = ( \bx, t)\int_{E_{g-1}}^{E_g} dE \, b(\bx,\bsOmega_m,E,t).
\end{equation}
Equation~\eqref{eq:transport-sn} is a set of $M \cdot G$ PDEs that must be further discretized in $x$ and $t$.  In this work, we use the diamond difference discretization in $x$ (see e.g., \cite[Section 3.3]{lewis1993}) and backward Euler time integration.
For one-dimensional spherical geometry, angular differencing coefficients, as described in \cite[Section 3.4]{lewis1993}, are used in addition to the diamond difference discretization. 
The use of implicit time integration is standard because of the time scales involved \cite{larsen2010advances}; various  higher-order temporal discretizations are also possible \cite{crockatt2017arbitrary,crockatt2019hybrid} but not critical for our presentation.  
Since diamond differencing is fairly standard, we omit the details here. 

The fully discrete approximation of the angular flux $\psi_{m,g}$ is computed using a nested iteration procedure that includes an outer loop for the energy groups and an inner iteration for the angles.  The procedure is stated in Algorithm~\ref{alg:multi} of Appendix~\ref{sec:algorithms}.
A standard sweeping method~\cite{bell1970nuclear} forms the core of the inner iteration.
In this process, the angular flux is calculated at the cell boundary and used to average the flux at the cell center through the diamond difference method.
The inner iteration will sweep in all angular directions using a fixed-point source iteration scheme until convergence of a scalar flux $\barpsi_{g}$ at a specific energy group. 
The convergence of a one group flux is used to update the outer iteration, which uses the Gauss-Seidel method over each energy group.
This method will converge quickly when there is minimal upscattering~\cite{lewis1993}.

\subsection{The hybrid multigroup, discrete ordinates equations}
\label{subsec:hybrid}
In the hybrid approach, the multigroup, discrete ordinates approximation is applied to the uncollided, collided, and total flux equations.   However, we now allow for the possibility that different levels of resolution are used in each case.  We will assume here that the uncollided and total flux equations are discretized at the same high resolution using $G$ groups and $M$ angles, while the collided equation uses lower resolution $\hat{G}$ and $\hat{M}$ angles.  To differentiate between different discretization parameters, a hat adornment $(\hat{\cdot})$ is used for the collided parameters.   Thus the weights and angles in the discretization of the collided equations are denoted by $\hw_{\hm}$ and $\hOmega_{\hm}$, respectively, for ${\hm} = 1, \dots, \hM$. To define the coarse groups, let
\begin{equation}
    0 = \Gamma_0 < \Gamma_1 \dots < \Gamma_{\hat{g}} \dots < \Gamma_{\hat{G}} =  G
\end{equation}
be a set of $\hat{G}+1$ integers
and set $\hE_{\hg} =  E_{\Gamma_{\hat g}}$.  Then for each $\hat{g} \in \chG$,
\begin{equation}
    \Delta \hE_{\hg} 
    = \hE_{\hg} - \hE_{\hg-1} 
    = E_{\Gamma_{\hat g}} - E_{\Gamma_{\hat g-1}}
    = \sum_{g=\Gamma_{\hg-1}+1}^{ \Gamma_{\hg}}
    E_{g} - E_{g-1}
    = \sum_{g=\Gamma_{\hg-1}+1}^{ \Gamma_{\hg}} \Delta E_g.
\end{equation}

The uncollided equation for 
\begin{equation}
    \psiu_{m,g} \approx \int_{E_{g-1}}^{E_{g}} dE \, \Psiu(\bx, \bsOmega_m, E, t)
\end{equation}
is
\begin{equation} \label{eq:sn-uncollided}
    \frac{1}{c_g} \frac{\partial }{\partial t} \psiu_{m, g} + \bsOmega_m \cdot \nabla \psiu_{m, g} + \sig[g]{t} \psiu_{m, g} = \qu_{m,g}, 
\end{equation} where $\qu_{m,g} = q_{m,g}$. 
We solve Eq. \eqref{eq:sn-uncollided} over a time-step $[t^n,t^{n+1})$ with initial condition 
\begin{equation}
    \psiu_{m, g} (\bx,t^n) = 
    \begin{cases}
    f_{m, g}(\bx), & \bx \in D, \quad m \in \cM, \quad g\in\cG, \quad t^n = 0 \\
    \psit_{m, g}(\bx,t^n_-), &  \bx \in D,\quad m \in \cM, \quad g\in\cG, \quad t^n > 0 
    \end{cases} ,
\end{equation} 
and incoming boundary data 
\begin{equation}
    \psiu_{m, g} = b_{m, g} (\bx, t) \qquad \text{for} \quad \bn(\bx) \cdot \bsOmega_m < 0, \quad m \in \cM, \quad g\in\cG\quand t>0.
\end{equation}
The collided transport equation for
\begin{equation}
    \psic_{\hat m,\hat g} \approx \int_{E_{\hat g-1}}^{E_{\hat{g}}} dE \, \Psic(\bx, \hat{\bsOmega}_m, E, t)
\end{equation}
is 
\begin{equation} \label{eq:sn-collided}
    \frac{1}{c_{\hg}} \frac{\partial }{\partial t} \psic_{\hm, \hg} + \hat \bsOmega_\hm \cdot \nabla \psic_{\hm, \hg} + \hsig[g]{t} \psic_{\hm, \hg} = 
    \sum_{g'=1}^{\hG} \hsig[\hg' \rightarrow \hg]{s} \barpsic_{\hg'}
    + \hat \chi_{\hg} \sum_{\hg'=1}^{\hG} \hat \nu_{\hg'} \hsig[\hg' \rightarrow \hg]{f} \barpsic_{\hg'} + \qc_{\hg},
\end{equation} 
where 
\begin{equation} \label{eq:coarsen-one}
c_{\hg} = \frac{1}{\Delta \hat{E}_{\hg}} \sum_{g=\Gamma_{\hg-1}}^{\Gamma_{\hg}} c_g, \quad
   \barpsic_{\hg'} = \sum_{\hm=1}^{\hM}  \hw_{\hm} \psic_{\hm,\hg'}
   ,\quad
   \hat \chi_{\hg} 
    = \sum_{g=\Gamma_{\hg}+1}^{\Gamma_{\hg+1}} \chi_{g}
   \quand
   \hat \nu_{\hg'} 
    = \sum_{g=\Gamma_{\hg}+1}^{\Gamma_{\hg+1}} \nu_{g};
\end{equation}
the energy-coarsened cross-sections are given by
\begin{equation} \label{eq:coarsen-two}
     \hsig[\hg]{t} = \frac{1}{\Delta \hat{E}_{\hg}} \sum_{g=\Gamma_{\hg-1}+1}^{\Gamma_{\hg}} \Delta E_{g} \sig[g]{t}
    \quand
        \hsig[\hg]{\ell} = \frac{1}{\Delta \hat{E}_{\hg}} \sum_{g=\Gamma_{\hg-1}+1}^{\Gamma_{\hg}} \sum_{g'=\Gamma_{\hg-1}+1}^{\Gamma_{\hg}} \Delta E_{g'} \sig[g'\rightarrow g]{\ell}, \quad \ell \in \rm\{s,f\} ;
\end{equation}
and the isotropic source is
\begin{equation} \label{eq:collided-source}
\qc_{\hg} = \sum_{g=\Gamma_{\hg-1}+1}^{\Gamma_{\hg}}  \sum_{g'=1}^{G} \sig[g' \rightarrow g]{s} \barpsiu_{g'}
+ \sum_{g=\Gamma_{\hg-1}+1}^{\Gamma_{\hg}} \chi_{g} \sum_{g'=1}^{G} \nu_{g'} \sig[g' \rightarrow g]{f} \barpsiu_{g'},
\end{equation}
with
\begin{equation}
    \barpsiu_{g} 
        = \sum_{m=1}^M w_m \psiu_{m,g} 
    \quand
    \barpsic_{\hat g} 
        = \sum_{\hat m=1}^{\hat M} \hw_{\hm} \psic_{\hat{m},\hat g}.
\end{equation}
We solve Eq. \eqref{eq:sn-collided} over a time-step $[t^n,t^{n+1})$ with initial condition 
\begin{equation}
    \psic_{\hm, \hg} (\bx,t^n) = 0 \qquad \text{for} \quad \bx \in D, \quad \hm \in \chM:=\{1,\cdots,\hM\}, \quad \hg\in\chG, \quad t^n \geq 0
\end{equation} 
and boundary data
\begin{equation}
    \psic_{\hm, \hg} = 0 \qquad \text{for} \quad \bn(\bx) \cdot \hOmega_{\hm} < 0, \quad \hm \in \chM, \quad \hg\in\chG, \quad t>0.
\end{equation}
The total flux equation for
\begin{equation}
\psit_{m,g}(x,t) \approx \int_{E_{g-1}}^{E_{g}} dE \, \Psit(\bx, \bsOmega_m, E, t)
\end{equation} 
is 
\begin{equation} \label{eq:sn-total}
    \frac{1}{c_g} \frac{\partial }{\partial t} \psit_{m, g} + \bsOmega_m \cdot \nabla \psit_{m, \hat g} + \sig[g]{t} \psit_{m, g} = \qt_{m,g}
\end{equation} 
where 
\begin{equation} \label{eq:transport-total-source}
    \qt_{m,g} = \frac{\Delta E_g}{\Delta \hE_{\hg}}
    \left[\qc_{\hg} +
    \sum_{\hg'=1}^{\hG} \sig[\hg' \rightarrow \hg]{s} \barpsic_{\hg'}
    + \chi_{\hg} \sum_{\hg'=1}^{\hG} \nu_{\hg'} \sig[\hg' \rightarrow \hg]{f} \barpsic_{\hg'} \right]
\end{equation}
and $\hg$ is the unique integer such that 
$\Gamma_{\hg-1} +1 \leq g \leq \Gamma_{\hg}$ or, equivalently, 
$(E_{g-1},E_{g}) \subset (\hE_{\hg-1},\hE_{\hg})$.
We solve Eq. \eqref{eq:sn-total} over a time-step $[t^n,t^{n+1})$  with initial data
\begin{equation}
    \psit_{m, g} (\bx,t^n) = 
        \begin{cases}
    f_{m, g}(\bx), & \bx \in D, \quad m \in \cM, \quad g\in\cG, \quad t^n = 0 \\
    \psit_{m, g}(\bx,t^n_-), &  \bx \in D,\quad m \in \cM, \quad g\in\cG, \quad t^n > 0 
    \end{cases} ,
\end{equation} 
and incoming boundary condition
\begin{equation}
    \psit_{m, g} = b_{m, g}(\bx, t) \qquad \text{for} \quad \bn(\bx) \cdot \bsOmega_m < 0, \quad m \in \cM, \quad g\in\cG\quand t>0.
\end{equation}

In summary, Eqs.~\eqref{eq:sn-uncollided}, \eqref{eq:sn-collided}, and \eqref{eq:sn-total} are solved in succession for each time step.  The solution of \eqref{eq:sn-total} at the end of the time step provides the initial condition needed by Eqs.~\eqref{eq:sn-uncollided} and \eqref{eq:sn-total} at the next time step, providing the mortice and tenon to consistently join the uncollided and collided calculations. The initial condition for \eqref{eq:sn-collided} is set to zero for each time step.    

As with the monolithic equations \eqref{eq:transport-sn}, we use backward Euler and diamond differencing for the temporal and spatial discretizations, respectively, of  Eqs.~\eqref{eq:sn-uncollided}, \eqref{eq:sn-collided}, and \eqref{eq:sn-total}.  Only one Euler step is used to advance the solutions from $t^n$ to $t^{n+1}$, and we use the same time steps and spatial mesh for the hybrid and monolithic equations.
In cases with time-dependent cross sections, the appropriate measures would be taken.  
For the purpose of this paper, the cross sections do not change in time.

The solution procedure for solving the collided flux is similar to the approach used in the monolithic case. It is described in Algorithm~\ref{alg:hybrid} of Appendix~\ref{sec:algorithms}.
It relies spatially on the very same sweeps and fixed-point source iteration for solving the flux at each energy group, while  Gauss-Seidel is used to integrate over groups~\cite{bell1970nuclear, lewis1993}.
The uncollided and the total flux updates do not require any iterations because the right-hand side of the respective equation is fixed.

\section{Numerical results} \label{sec:results}

In this section, we demonstrate the computational benefits of the hybrid using several time-dependent problems with different one-dimensional geometries. 
While these examples use isotropic scattering, we expect to see similar benefits with anisotropic scattering despite the coupling terms being modified. This is an area for future research.
It should also be noted that the cross sections do not change in time for these results.

In the slab case, the reduction of the equations above is straightforward. 
In the spherical case, the coordinate system introduces angular derivatives that must be discretized.  We do so using the standard approach given in \cite[Section 3.4]{lewis1993}.
Our reference solution uses the original number of energy groups as constructed by the Fudge~\cite{fudge} software package for nuclear data management, but the groups are coarsened according to the formulas in \eqref{eq:coarsen-two}.

Three different solution strategies are used for each test problem:
\begin{itemize}
    \item {\bf Multigroup.} This is the traditional neutron multigroup method with $\hG$ groups and $\hM$ angles. 
    It uses Gauss-Seidel with a tolerance of $\varepsilon_G = 1 \times 10^{-12}$ for the outer iteration and source iteration with a tolerance of $\varepsilon_M = 1 \times 10^{-8}$ for the inner iteration, as described in Algorithm~\ref{alg:multi}.  
    The coarsening strategy in Eqs.~\eqref{eq:coarsen-one} and~\eqref{eq:coarsen-two} is used for lower fidelity multigroup models.  The accuracy of the multigroup method is not necessarily monotonic in the number of groups.  In particular, it can happen that smaller number of groups yield better answers due to the nonlinearity of the procedure, i.e., if the solution is separable in energy, space, and angle, a single group calculation can be exact \cite[\S 4.3]{bell1970nuclear}. Additionally, there can be cancellation of errors in integrated quantities.  This  behavior is observed in some of the test cases below.
    
    \item {\bf Hybrid.} This is the hybrid method described in Section \ref{subsec:hybrid}.  It uses $G$ groups and $M$ angles in the discretization of the total and uncollided angular fluxes, and it uses $\hG$ groups and $\hM$ angles in the discretization of the collided angular flux.
    It uses the algorithm described in Algorithm~\ref{alg:hybrid} with iteration tolerances of $\varepsilon_G = 1 \times 10^{-12}$ and $\varepsilon_M = 1 \times 10^{-8}$ for the outer and inner iterations, respectively.
    
   \item {\bf Splitting.}  This method is the same as the hybrid method described in  Algorithm~\ref{alg:hybrid} with the same iteration tolerances ($\varepsilon_G = 1 \times 10^{-12}$ and $\varepsilon_M = 1 \times 10^{-8}$), but it uses $\hG$ groups and $\hM$ angles for the collided, uncollided, and total flux. 
   It is included to investigate the effects of the hybrid discretization vs. the solver strategy: on one hand, it should produce the same answers (up to iteration tolerances) as the multigroup method but may show different convergence behavior because the solver strategy is different.
   On the other hand, it should be less expensive, but also less accurate than the hybrid that uses finer resolution for the uncollided and total fluxes.  
\end{itemize}

\begin{figure}[!ht]
    \centering
    \subfloat[]{\includegraphics[width=0.75\textwidth]{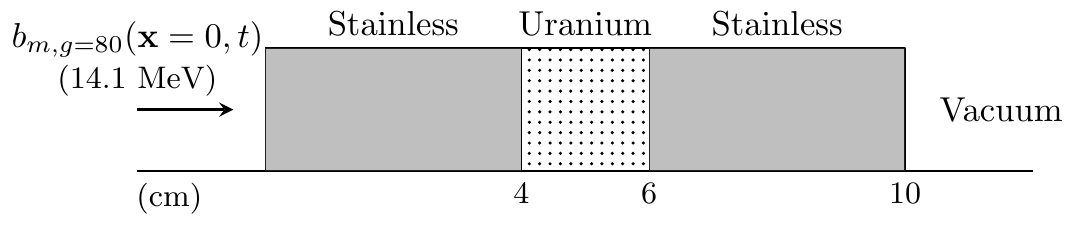}\label{fig:uranium-slab-setup}} \\
    \subfloat[]{\includegraphics[width=0.75\textwidth]{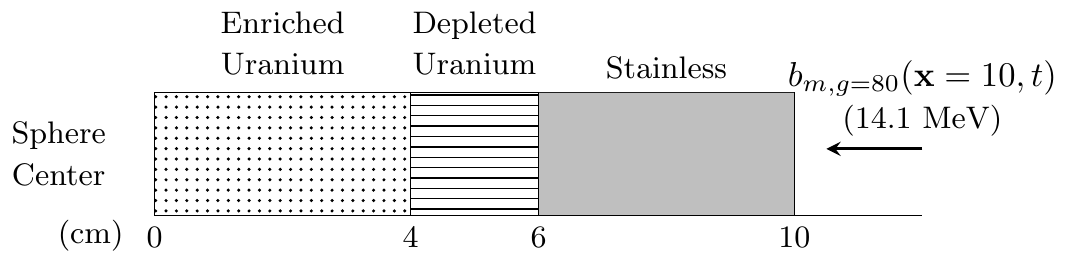}\label{fig:uranium-sphere-setup}}
    \caption{General setup of the (a) enriched uranium slab and the (b) enriched and depleted uranium sphere. The enriched uranium in both of these problems uses 20\% enriched uranium-235. The depleted uranium used only uranium-238. The boundary source performs in the same manner for both of these problems, which is initially at unity and decays by a half every 0.1 $\mu$s after the initial 0.2 $\mu$s, with the elapsed time set at 1 $\mu$s, as described in Eq.~\eqref{eq:decay-source}.}
    \label{fig:uranium-setup}
\end{figure}

To compare the accuracy between the different models, the total fission rate density (FRD) and the fission rate root mean squared error (RMSE) are used. 
The total fission rate density is the sum of the fission rate at each spatial cell, or more precisely, \begin{equation} \label{eq:fission-rate-density}
    {\rm{FRD}} = \; \sum_{g=1}^{G} \left (\chi_{g} \sum_{g'=1}^{G} \nu_{g'} \sig[g' \rightarrow g]{f} \barpsi_{g'} \right),
\end{equation}
with units of cm$^{-3} $s$^{-1}$. 
To compare the wall clock times between the different models, we compute
\begin{equation} \label{eq:wall-clock-diff}
    \tau' = \frac{\tau_{\rm{mg}} - \tau_{\rm{hy}}}{\tau_{\rm{mg}}}, 
\end{equation} where $\tau_{\rm{mg}}$ and $\tau_{\rm{hy}}$ are the wall clock times required to run the multigroup and hybrid simulations, respectively. 
This format was used to show that positive wall clock time differences are for faster hybrid method simulation times. 
To account for both the wall clock time and accuracy in one metric, we use a modified figure of merit (FOM) that is commonly used in the Monte Carlo community \cite{figureofmerit}:
\begin{equation} \label{eq:fom}
    \FOM = \frac{1}{\varepsilon \, \tau},
\end{equation} 
where $\varepsilon$ is the RMSE  and $\tau$ is the wall clock time\footnote{For Monte Carlo, $\varepsilon^2$ is used in place of $\varepsilon$ in Eq.~\eqref{eq:fom} as it is the variance in a statistical estimate.}.
A larger figure of merit indicates a more efficient calculation.

\begin{figure}[!ht]
    \centering
    \subfloat[]{\includegraphics[width=0.48\textwidth]{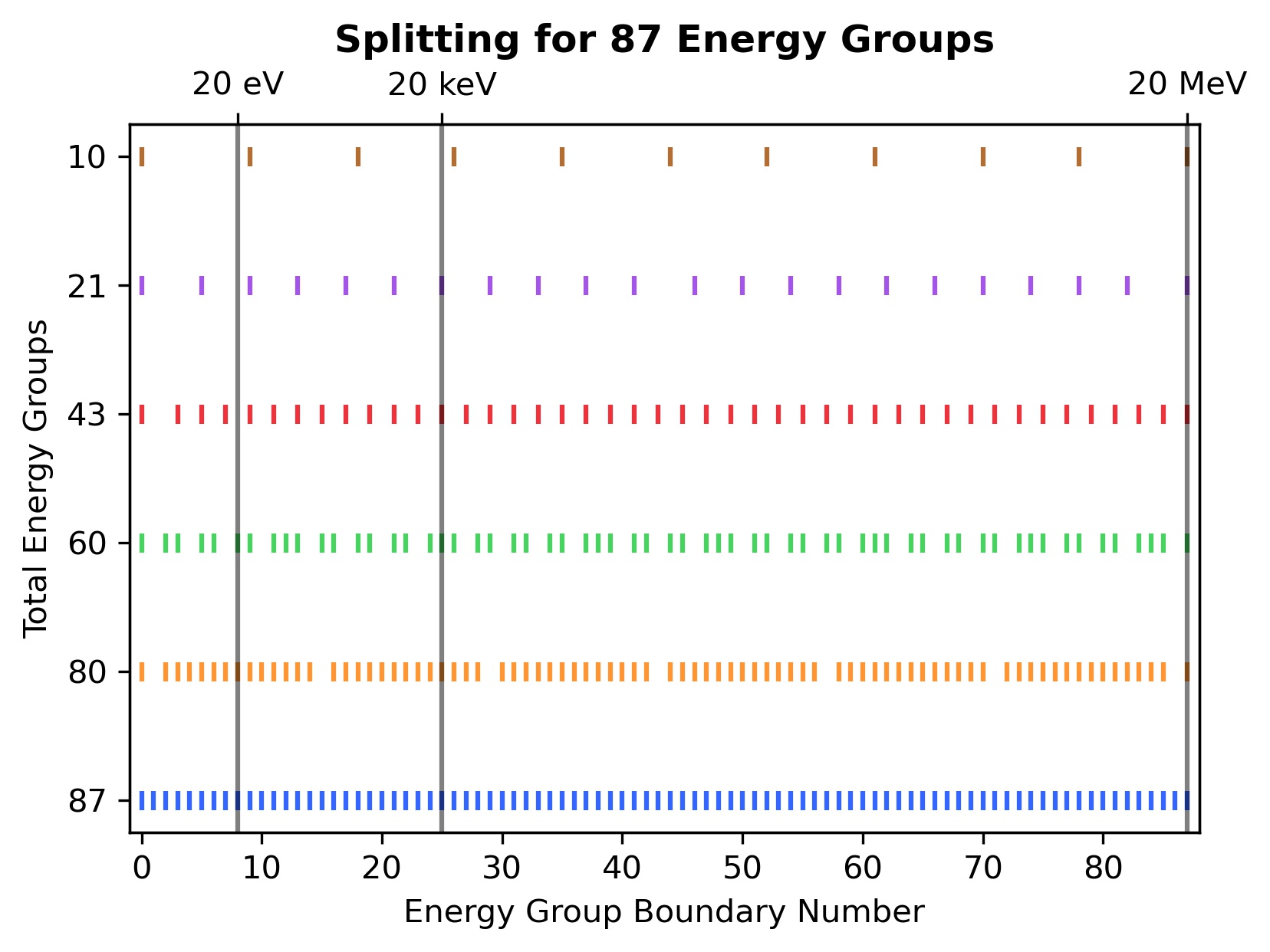}\label{fig:energy-splitting-087}} 
    \subfloat[]{\includegraphics[width=0.48\textwidth]{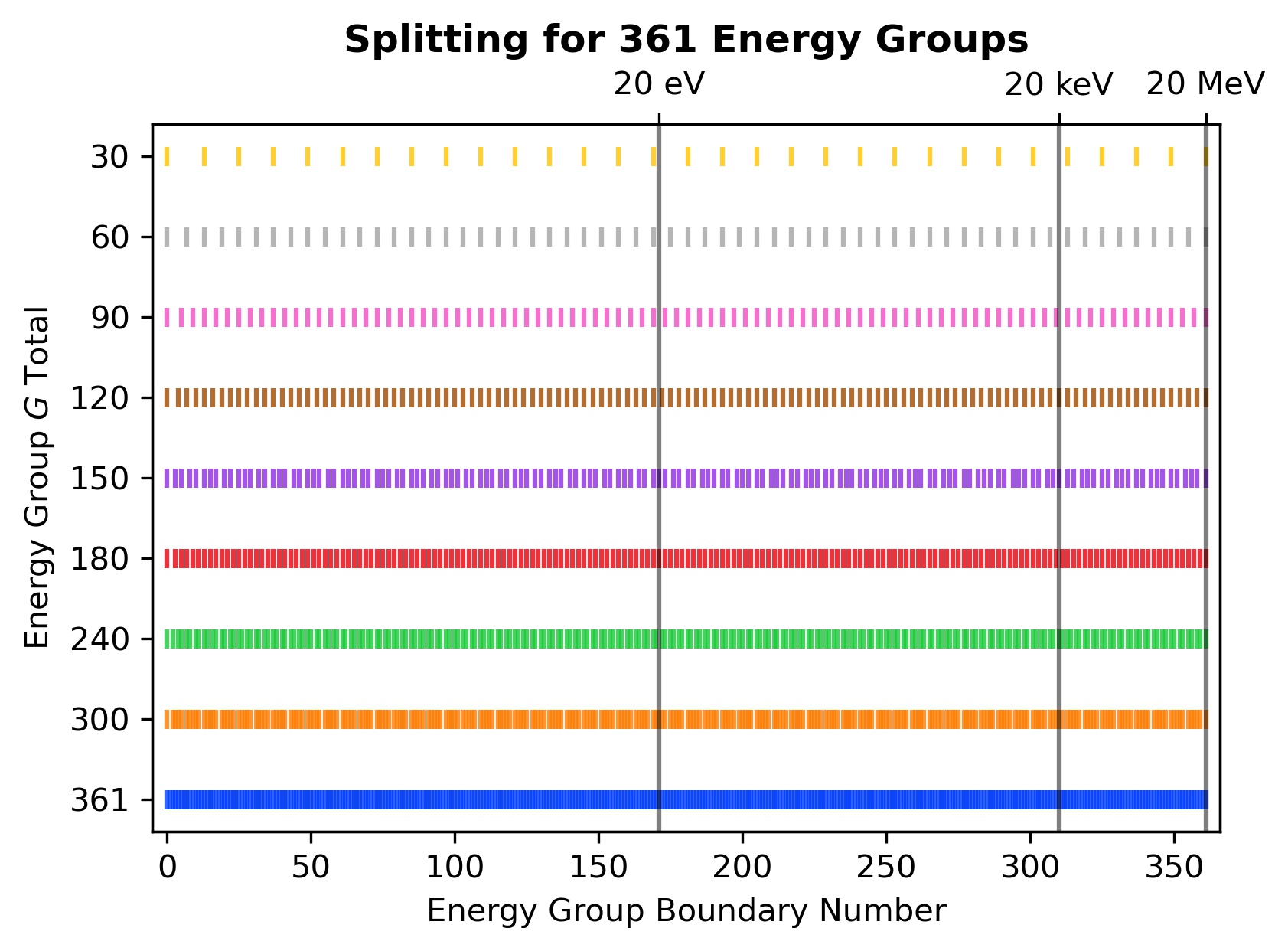}\label{fig:energy-splitting-361}}
    \caption{The coarsening results of the (a) $G = 87$ and (b) $G = 361$ energy grids as shown by the energy bounds. The naive coarsening of (a) can be seen with the coarser grids ($\hG = 10$, $21$) and how the neighboring energy groups of $\hG = 21$ are combined to form the bounds of the $\hG = 10$ energy bounds. The involved approach with (b) combines the energy groups based off the accuracy of the $k$-eigenvalue of a one-dimensional critical slab problem.}
    \label{fig:energy-splitting}
\end{figure} 

\subsection{Enriched uranium slab}
The first example problem uses a 10 cm slab with a final time of $T$ = 1 $\mu$s and a time step of $h$ = 0.01 $\mu$s.
Two different materials, 20\% enriched uranium and stainless steel, were used in the layout shown in Fig.~\ref{fig:uranium-slab-setup}.
An inflow condition for the 14.1 MeV, $g = 80$, energy group is prescribed at $\bx = 0$.  The inflow starts at one and decreases by a factor of two at prescribed intervals.  More specifically, 
\begin{equation} \label{eq:decay-source}
    b_{m, g=80} (\bx = 0, t) = 
    \begin{cases}
    1 ,&\ t \in [0,0.2] \\
    0.5^{k(t)}\left(1+ 2 \, \erfc \left( \frac{t-0.1(1+k(t))}{0.01} \right) \right) ,& \ t \in (0.2,1.0]
    \end{cases},
\quad
\text{where}
\quad
    k(t) = \left \lceil \frac{t-0.2}{0.1} \right \rceil
\end{equation}
and $t$ has units of $\mu$s.

Each drop in the input boundary induces a transient solution that eventually returns to a quasi steady state. 
Because it is expected that the largest numerical errors occur during these transient phases and for this reason, we focus on results examine $t$ = 0.21 $\mu$s, just after the input condition drops for the first time. These results are representative of the performance of the methods throughout the simulation. The wall clock time is calculated for the entire simulation and is repeated five times to obtain an average value.

The original energy grid for the uranium and stainless steel used $G = 87$ energy groups. 
The group coarsening is accomplished by taking a naive approach, that assigns the same number of fine energy groups to each coarse group, modulo differences in rounding.
While it is possible to optimize the coarsening procedure using the same techniques that were used to generate the reference energy grid, determining the optimal group structure was not the purpose of this research.
The results of the coarsening strategy are depicted in Fig.~\ref{fig:energy-splitting-087}. 

\begin{figure}[!ht]
    \centering
    \subfloat[]{\includegraphics[width=0.48\textwidth]{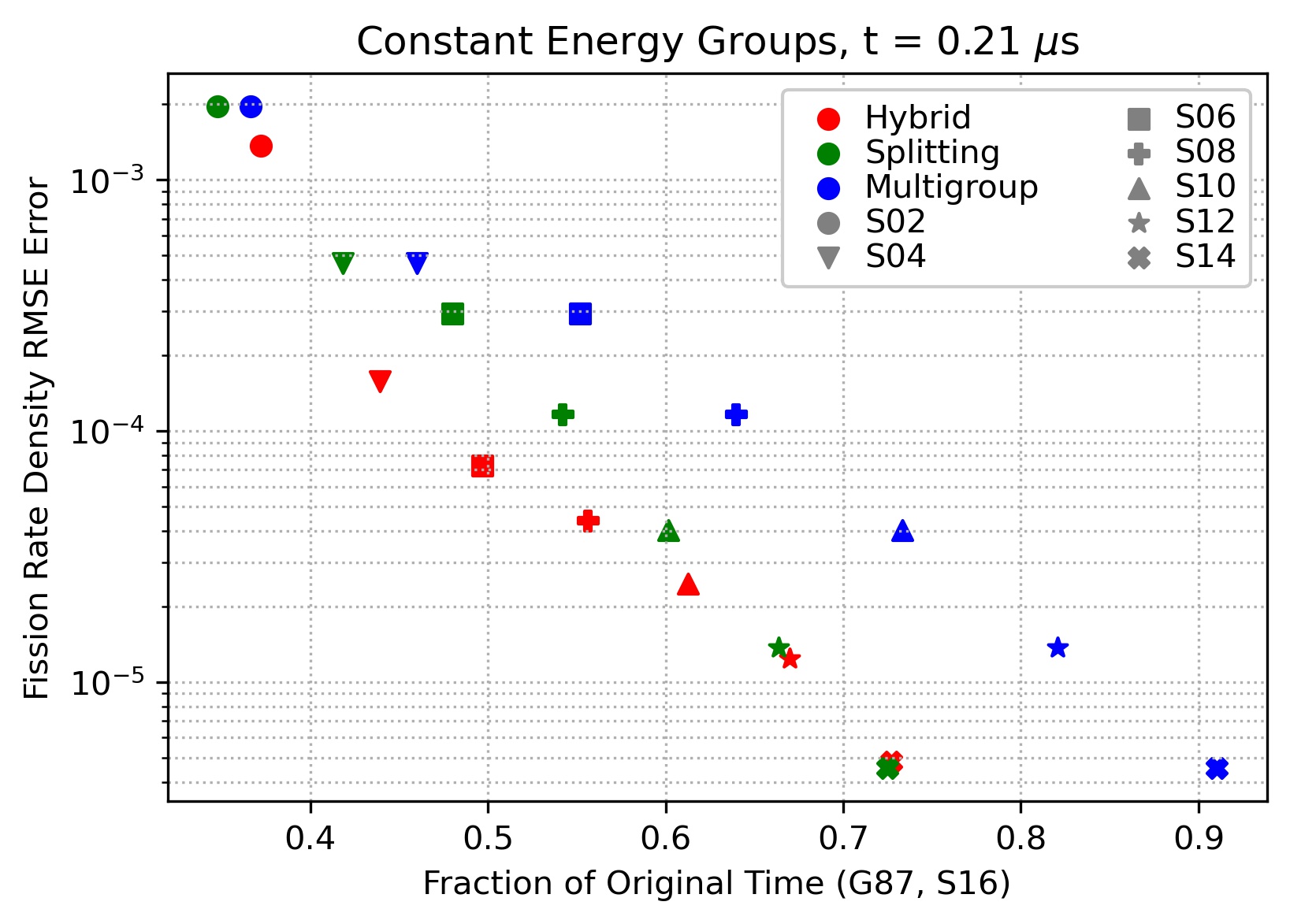} \label{fig:slab-constant-groups}}
    \subfloat[]{\includegraphics[width=0.48\textwidth]{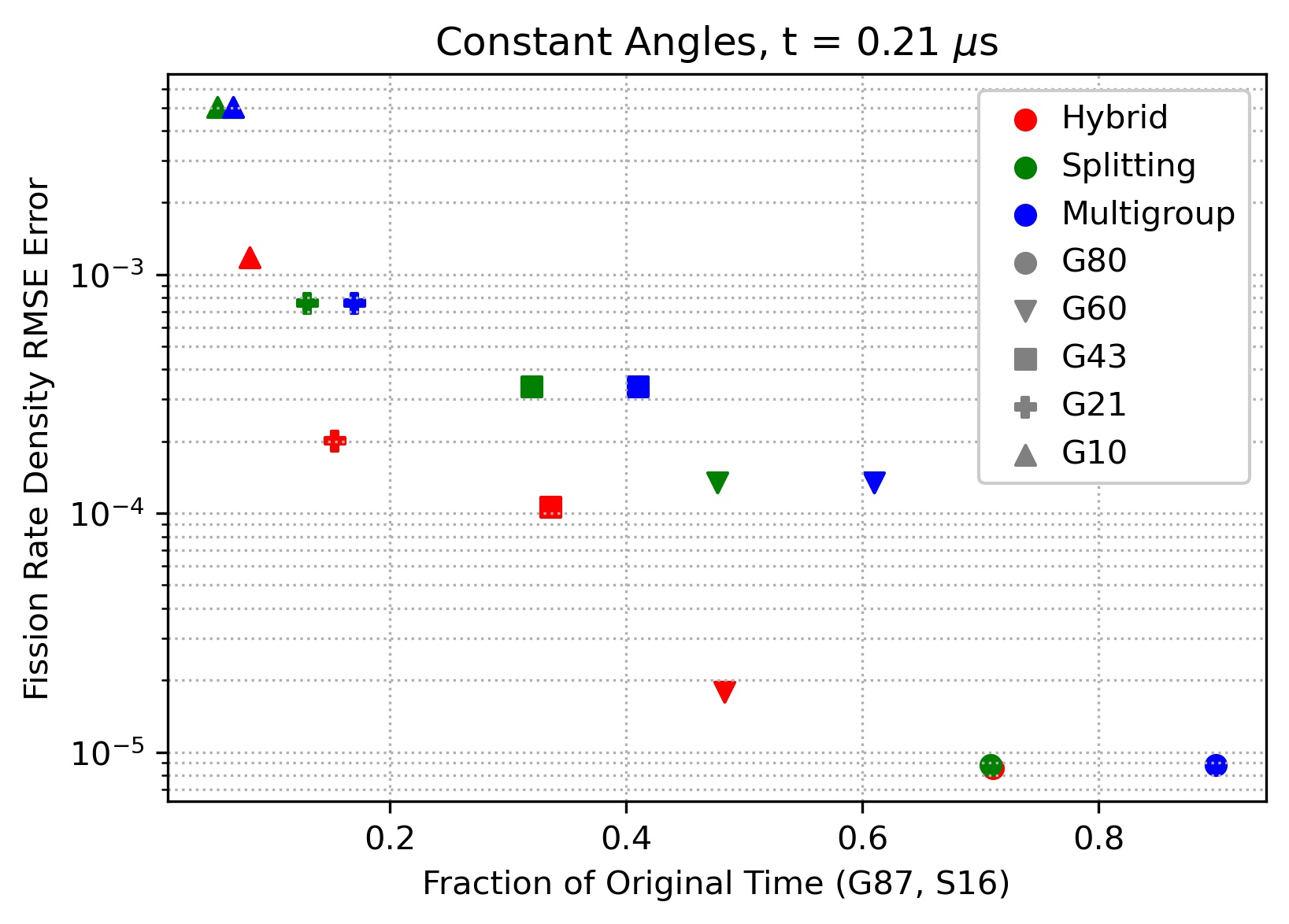} \label{fig:slab-constant-angles}}
    \caption{Comparison of the RMSE error in the fission rate density vs. wall clock time for the enriched uranium slab problem at time $t=0.21~\mu$s: (a) constant numbers of energy groups with varying number of angles and (b) constant number of angles with varying number of energy groups. As expected, the splitting method has the same error as the multigroup method, but takes less time to compute, with the difference in wall clock time increasing as $\hG$ and $\hM$ increase. The hybrid method is typically more accurate, but slightly more expensive than the splitting method.  As expected, these differences disappear as $\hG$ and $\hM$ increase.}
    \label{fig:slab-analysis}
\end{figure}

In Fig.~\ref{fig:slab-analysis}, we compare the error versus wall clock time of the three methods described at the beginning of the section, fixing either the number of groups or the number of angles.  
The reference solution for these comparisons is a grid with $G = 87$ energy groups and $M = 16$ angles. 
In Fig.~\ref{fig:slab-constant-groups}, we vary the number of coarse angles $\hat{M} < M = 16$ while fixing $\hG = G = 87$ energy groups.  
In Fig.~\ref{fig:slab-constant-angles}, we vary the number of coarse groups $\hat{G} < G = 87$, while fixing $\hM = M = 16$ angles. 
As expected, the multigroup and splitting results yield the same errors. 
However, the splitting runs faster and, moreover, the relative difference in computational time increases as $\hat{M}$ or $\hat{G}$ increases.  The improvement in the splitting comes from the different solver strategy.  In particular, the uncollided equation in the hybrid does not require inner iterations over angle to be converge, thereby avoiding to some degree the over-solving phenomena that can unnecessarily slow down convergence~\cite{senecal2017}.
This is shown in the full multigroup and hybrid models ($G = \hG = 87$, $M = \hM = 16$), in which the hybrid method required about 161 fewer iterations per time step.
The comparison between the hybrid and splitting methods show that the hybrid generally provides improved accuracy with a marginal cost increase.
The efficiency gains from the hybrid are substantial for intermediate values of $\hat{M}$ and $\hat{G}$, but it tend to zero as $\hM \to M$ or as $\hG \to G$, as expected since the methods are essentially the same in these limits. 

\begin{figure}[!ht]
    \centering
    \subfloat[]{\includegraphics[width=0.52\textwidth]{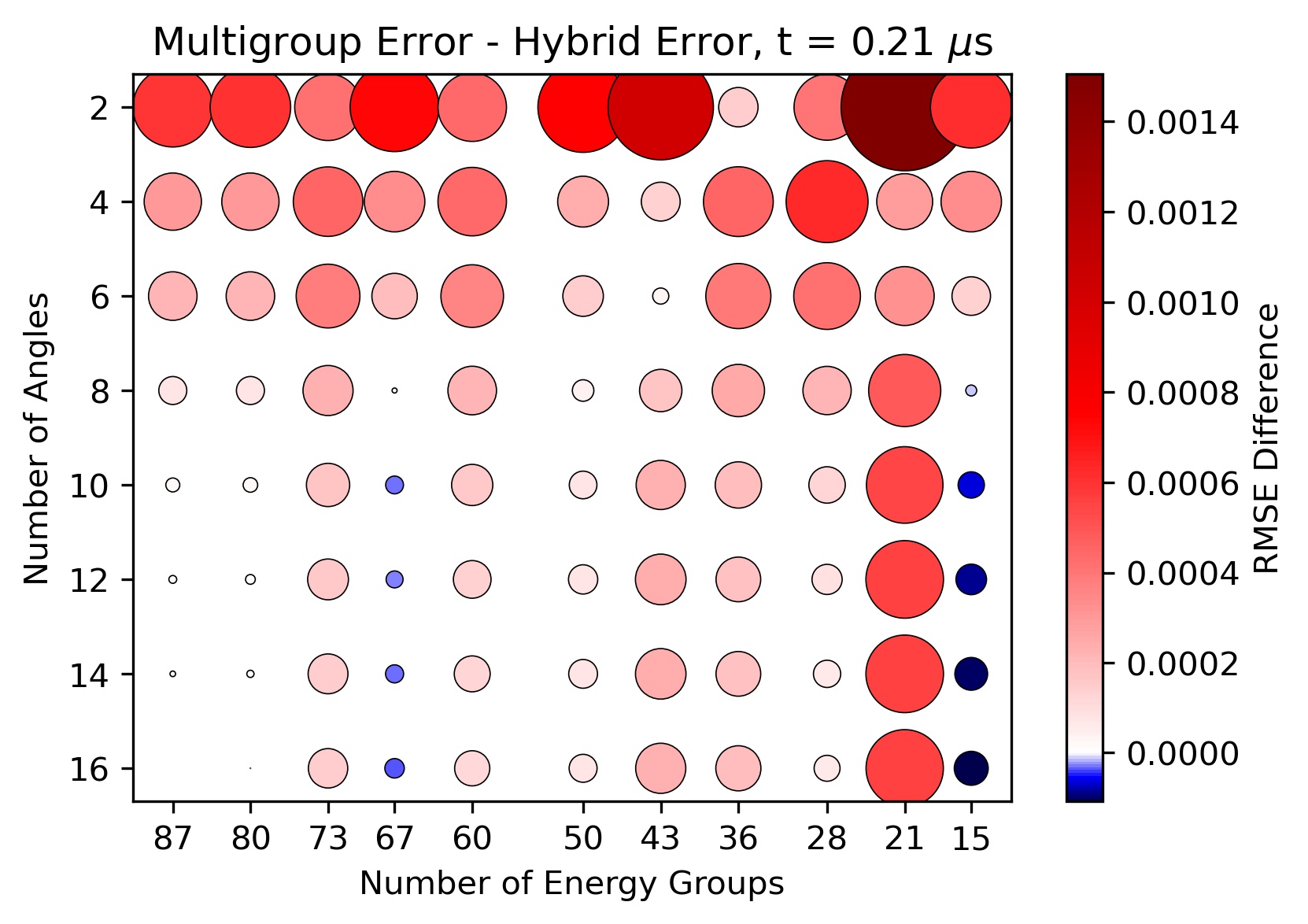} \label{fig:slab-error-grid}} 
    \subfloat[]{\includegraphics[width=0.52\textwidth]{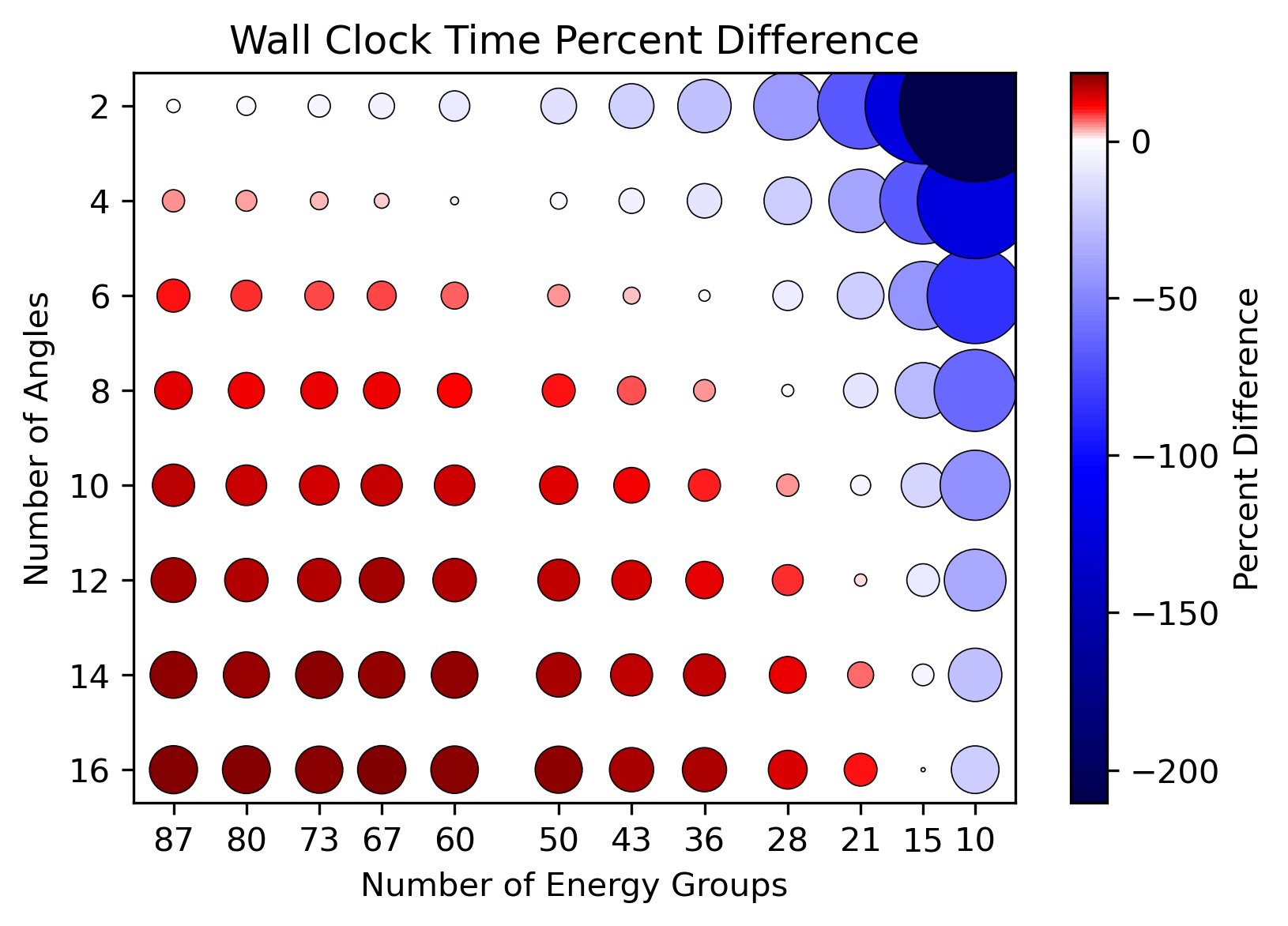} \label{fig:slab-time-grid}} \\
    \subfloat[]{\includegraphics[width=0.52\textwidth]{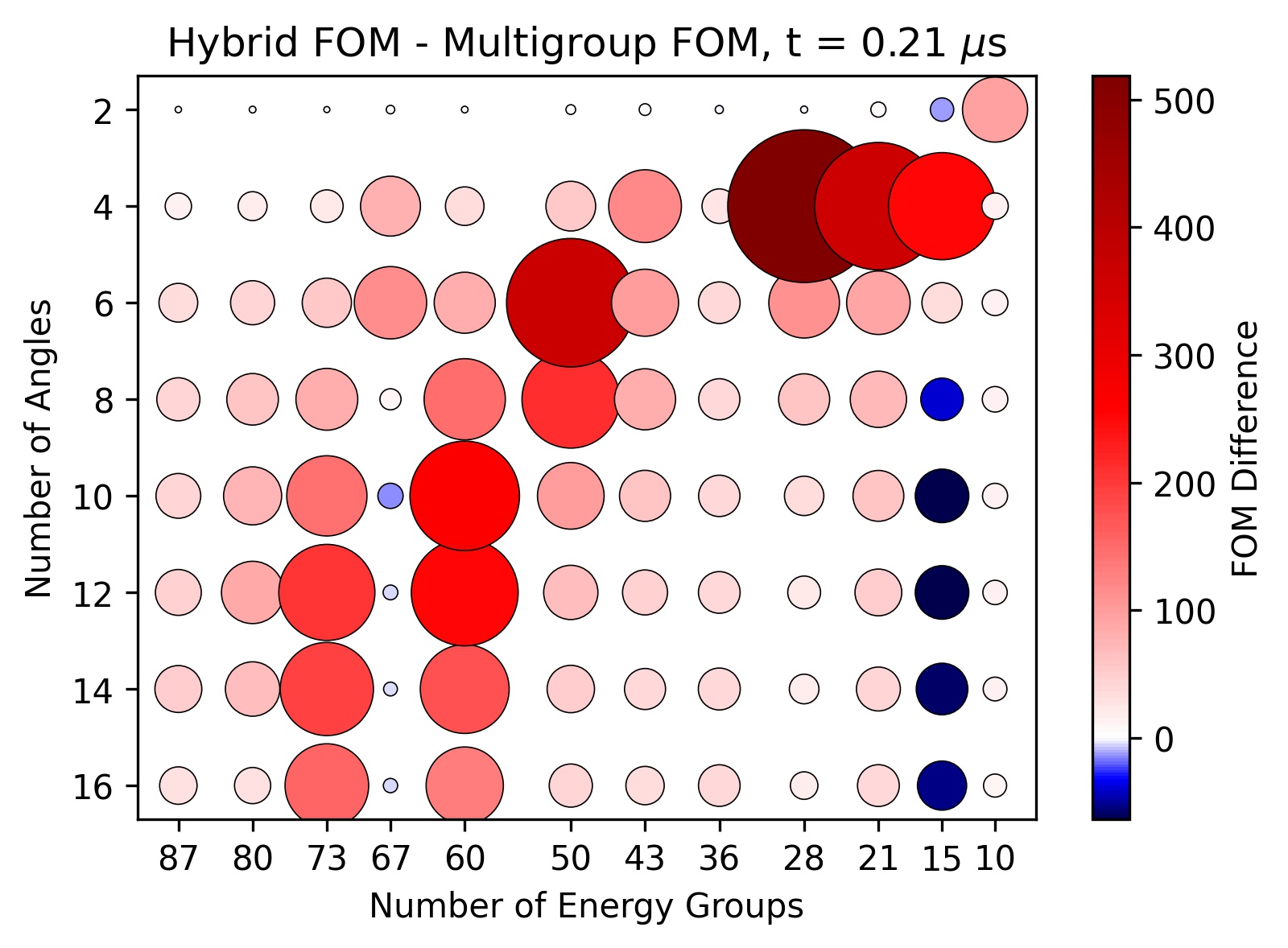} \label{fig:slab-fom}}
    \caption{Efficiency comparison between hybrid and multigroup methods applied to the enriched uranium slab problem across a range of discretization parameters: (a) multigroup error minus hybrid error; (b) percent difference in wall clock time; (c) hybrid FOM minus multigroup FOM.  In (a), errors are computed at $t$ = 0.21 $\mu$s with respect to a reference solution with $M = 32$ angles and $G = 87$ groups; a positive difference favors the hybrid. The hybrid models with higher error, $\hG = 67$ and $\hG = 15$ are most likely due to selection of the coarse groups.  The errors for $\hG = 10$ are quite large and have thus been omitted to make the remaining data easier to view. For (b), the percent difference in wall clock time is $(\tau_{\rm{mg}} - \tau_{\rm{hy}}) / \tau_{\rm{mg}}$; positive values mean that the hybrid is faster. In (c), the FOM difference is computed using \eqref{eq:fom}; positive values favor the hybrid.}
    \label{fig:slab-grid}
\end{figure}

In Fig.~\ref{fig:slab-grid}, we compare the efficiency of the multigroup and hybrid methods in computing the fission rate density over a range of energy and angle discretization parameters.
The reference solution for these comparisons employs a grid with $G = 87$ energy groups and $M = 32$ angles.  In Fig.~\ref{fig:slab-error-grid}, we show the differences in errors of the two methods.
For the majority of parameters, the hybrid method performs better, with the larger differences appearing in the coarser energy grids, but there are exceptions for larger values of $\hM$ when $\hG = 67$ and $\hG = 15$. 
These exceptions occur when the naive coarsening strategy for the multigroup method performs unexpectedly well. 
In Fig.~\ref{fig:slab-time-grid}, we show the differences in wall clock time for the same set of simulations.  Roughly speaking, we observe that the hybrid is faster for high-resolution calculations and the multigroup is faster for low-resolution calculations.
In Fig.~\ref{fig:slab-fom}, we plot the difference between the FOM for the hybrid method and the FOM for  multigroup method.
In the large majority of cases, the hybrid FOM is better. Exceptions to this trend occur when $\hG=15$ or $\hG=67$.

\begin{figure}[!ht]
    \centering
    \subfloat[]{\includegraphics[width=0.48\textwidth]{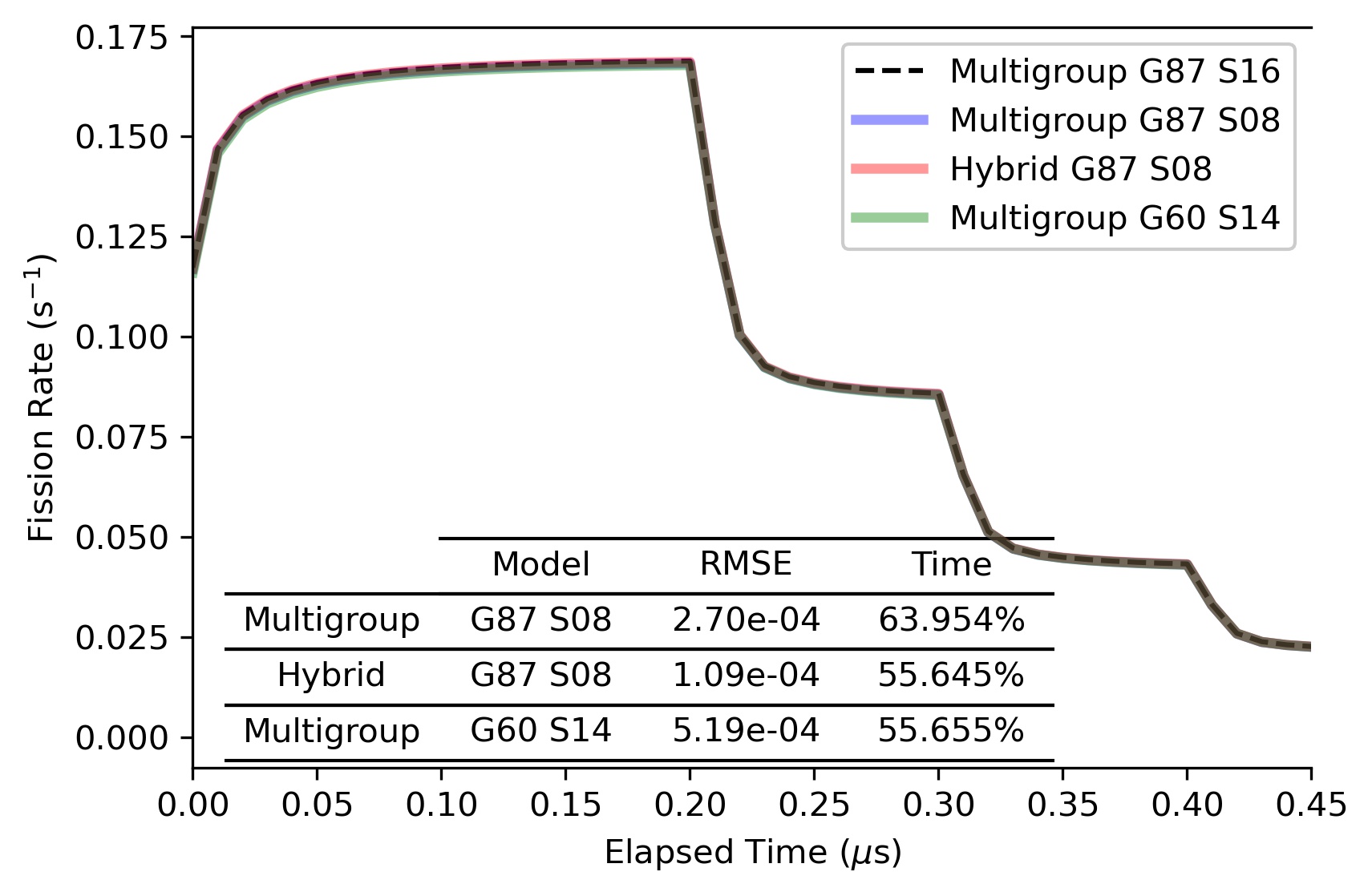} \label{fig:slab-fission-rate-time}}
    \subfloat[]{\includegraphics[width=0.48\textwidth]{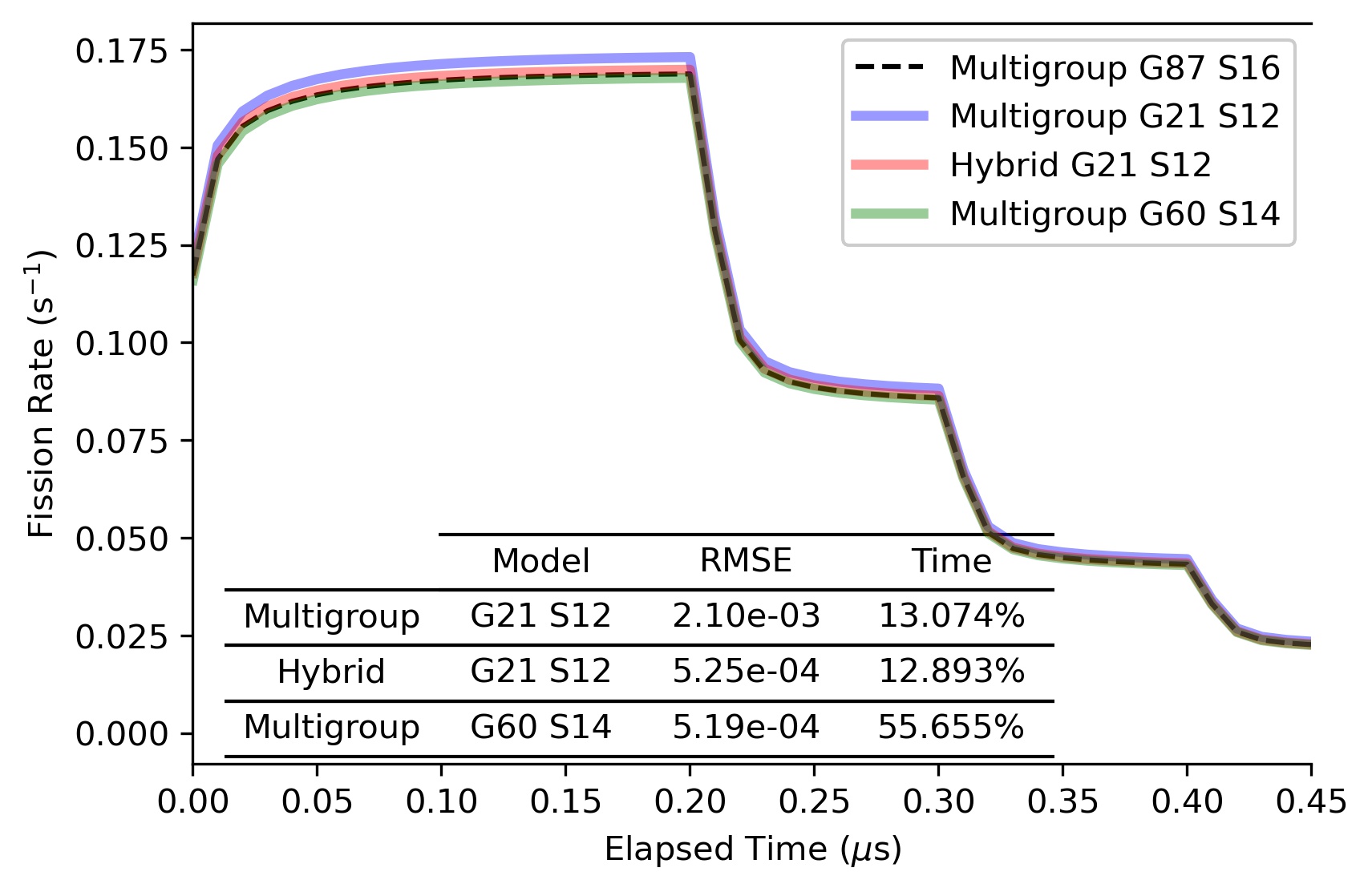} \label{fig:slab-fission-rate-error}}
    \caption{Illustrative comparisons of the hybrid and multigroup methods in computing the fission rate for the enriched uranium slab problem.  For similar wall clock times, the hybrid yields smaller errors.  For similar errors, the hybrid yields smaller wall clock times.}
    \label{fig:slab-fission-rate}
\end{figure}

In Fig.~\ref{fig:slab-fission-rate}, we plot illustrative results from the hybrid and multigroup calculations. 
These results demonstrate that for models with similar errors, the hybrid method yields faster computational times and for models with  comparable wall clock times, the hybrid method yields more accurate results.

\subsection{Enriched and depleted uranium sphere}
The second test problem involves a sphere of radius 10 cm made of three materials: stainless steel, 20\% $^{235}$U enriched uranium, and depleted uranium containing only uranium-238. A diagram of the setup is given in Fig.~\ref{fig:uranium-sphere-setup}.  The geometry of the problem allows it to to be modeled using a single spatial dimension for the radial direction.
The simulation is run to a final time $T$ = 1 $\mu$s with time steps of size $h$  = 0.01 $\mu$s.
The boundary condition used for the enriched uranium slab problem in Eq.~\eqref{eq:decay-source} was also used in this example, with the same 14.1 MeV energy group and the same decay rate, except that it is at the sphere edge, $b_{m,g}(\bx = 10, \, t)$. 

\begin{figure}[!ht]
    \centering
    \subfloat[]{\includegraphics[width=0.48\textwidth]{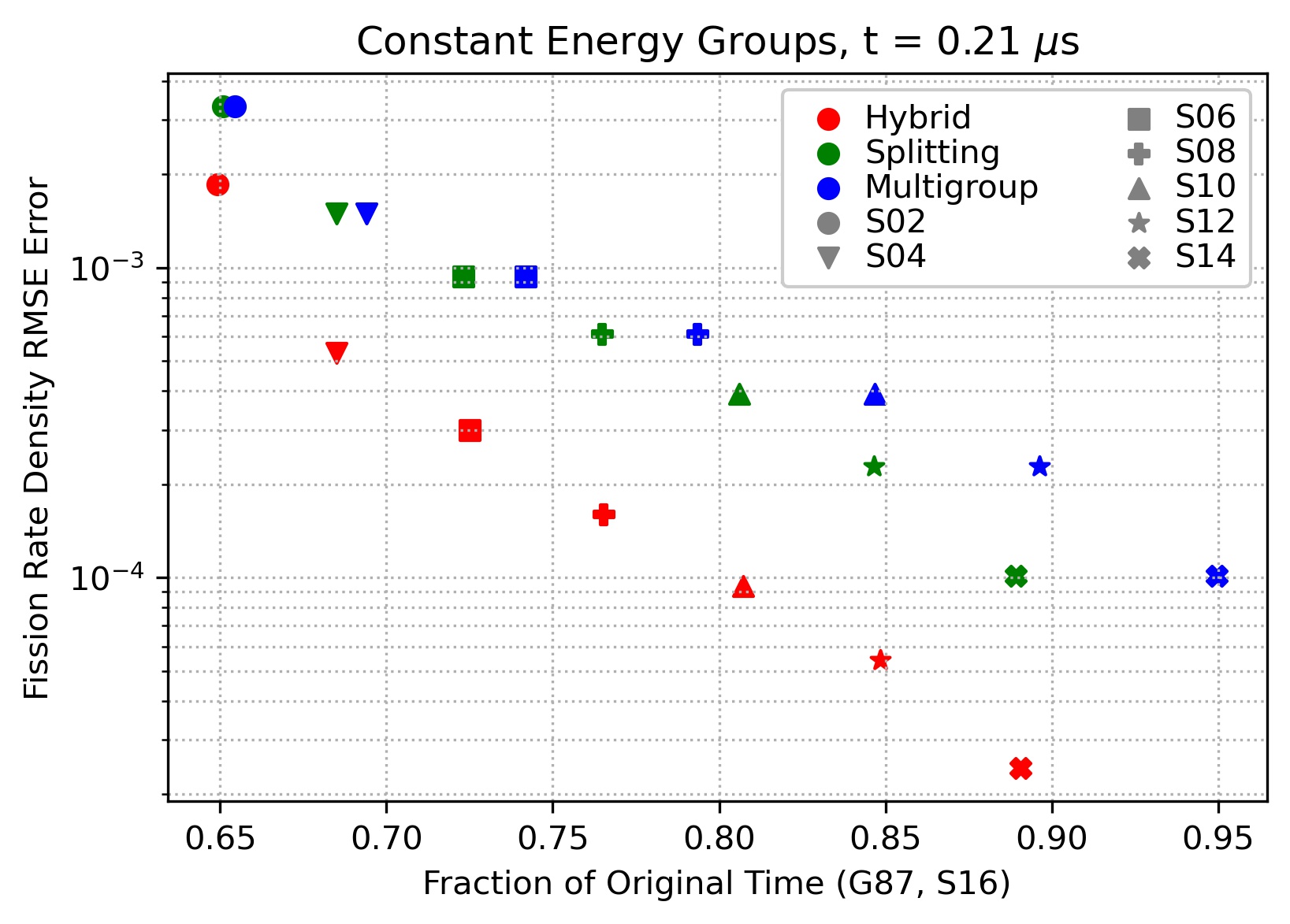} \label{fig:sphere-constant-groups}}
    \subfloat[]{\includegraphics[width=0.48\textwidth]{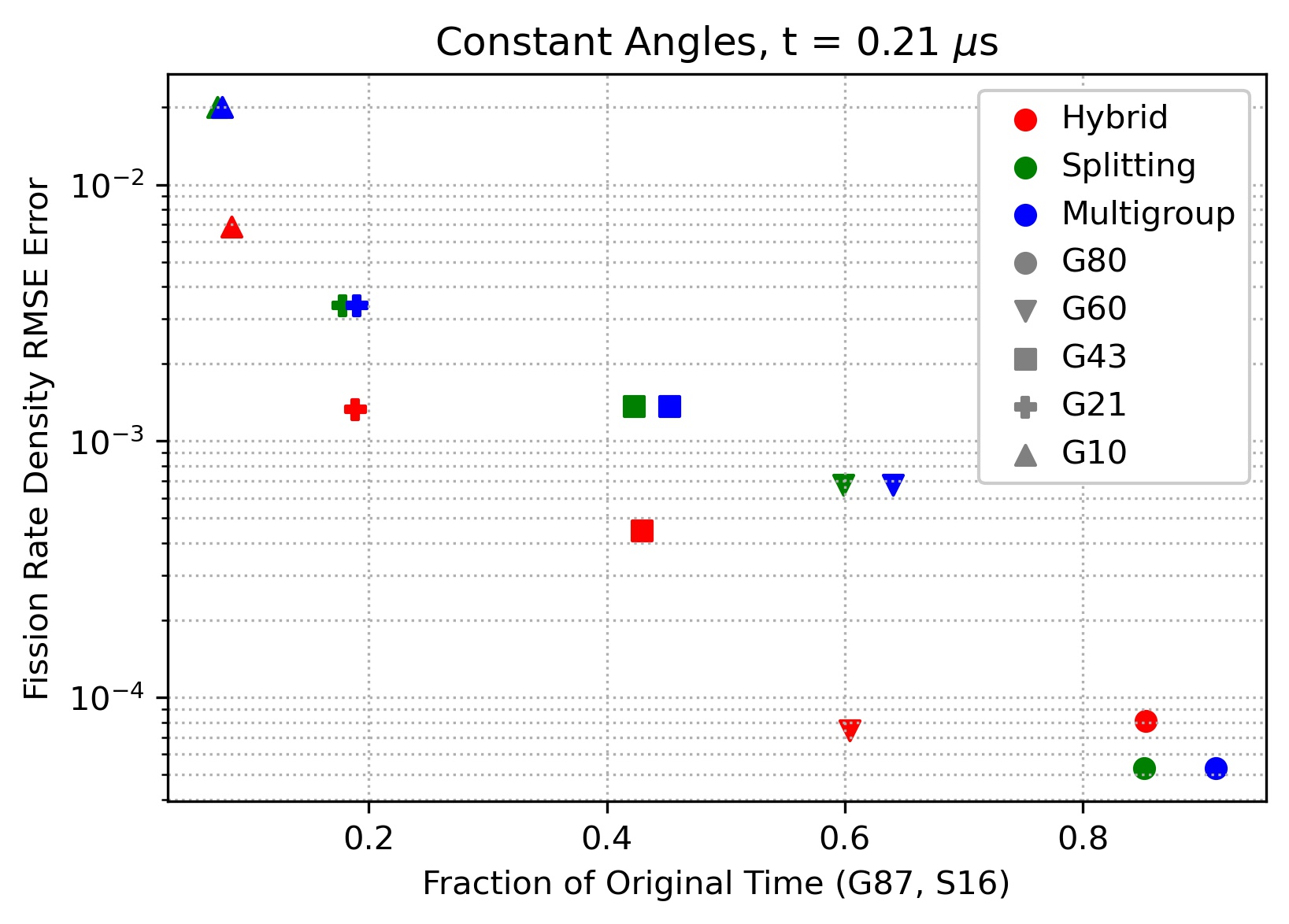} \label{fig:sphere-constant-angles}}
    \caption{Comparison  the RMSE error vs. wall clock time in the fission rate density for the enriched and depleted uranium sphere problem: (a) constant numbers of energy groups and varying number angles and (b) constant number of angles and varying number of energy groups. As expected, the splitting method has the same error as the multigroup method, but takes less time to compute, with the difference in wall clock time increasing as $\hG$ and $\hM$ increase. The hybrid method is typically more accurate, but slightly more expensive than the splitting method.}
    \label{fig:sphere-analysis}
\end{figure}

In Fig.~\ref{fig:sphere-analysis}, we compare the error versus the wall time for the multigroup, splitting, and hybrid methods, fixing either the number of groups or the number of angles. 
The reference solution for these comparisons is a grid with $G = 87$ energy groups and $M = 16$ angles. The coarse energy groups are the same as the  previous problem and are depicted in Fig.~\ref{fig:energy-splitting-087}.
In Fig.~\ref{fig:sphere-constant-groups}, we vary the number of coarse angles $\hat{M} < M = 16$ while fixing $\hG = G = 87$ energy groups. 
In Fig.~\ref{fig:sphere-constant-angles}, we vary the number of coarse groups $\hat{G} < G = 87$, while fixing $\hM = M = 16$ angles. 

Overall, we observe similar results as for the slab problem, although the improvements in wall clock time for the splitting method over the multigroup method are smaller and the improvements in accuracy of the hybrid over the splitting method are greater. 
As in the previous problem, the multigroup method may perform unexpectedly well in some situations because of the coarsening.  
For the current problem, this situation occurs when $\hG = 80$ and in this case, the multigroup method outperforms the hybrid.

\begin{figure}[!ht]
    \centering 
    \subfloat[]{\includegraphics[width=0.52\textwidth]{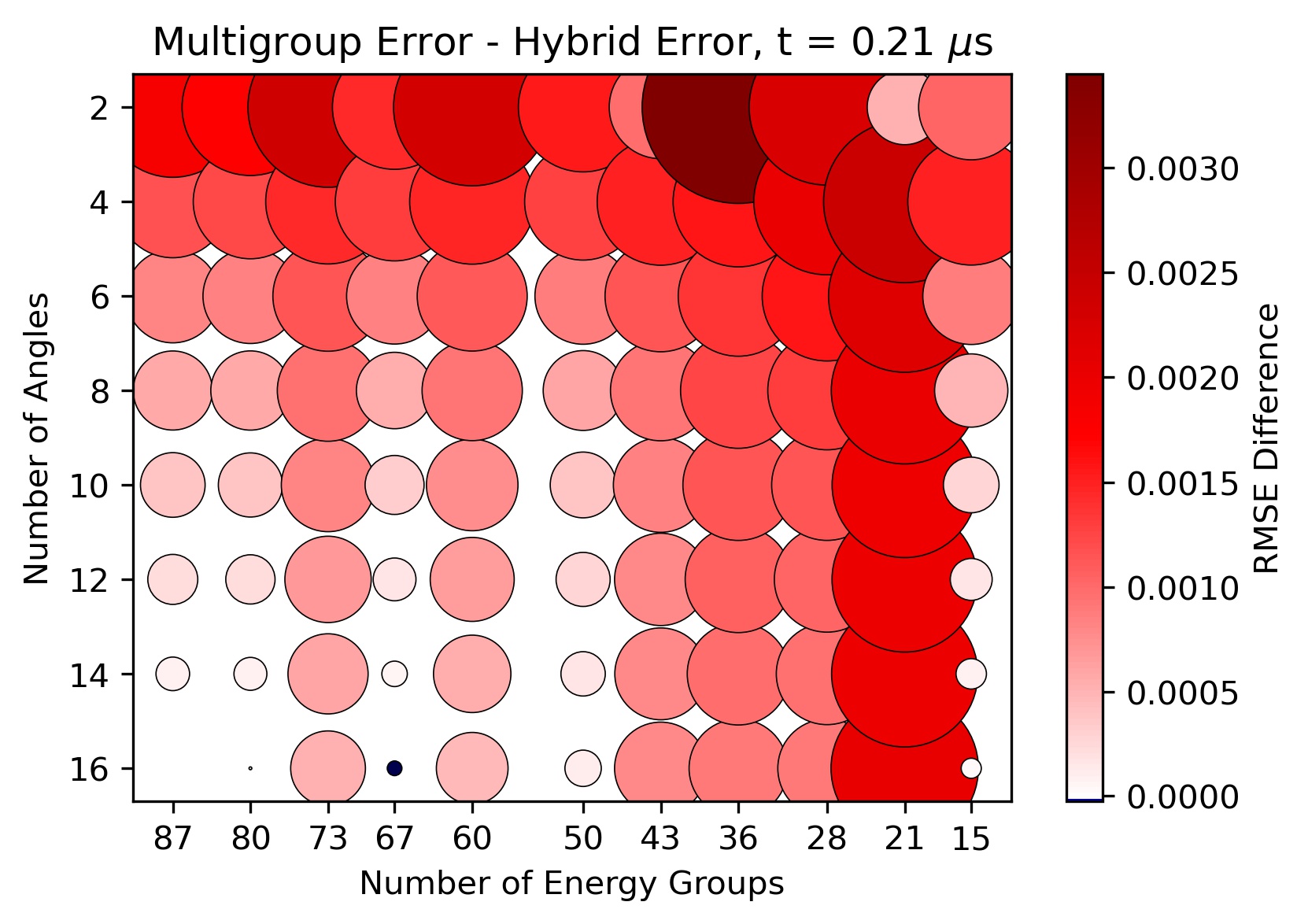} \label{fig:sphere-error-grid}} 
    \subfloat[]{\includegraphics[width=0.52\textwidth]{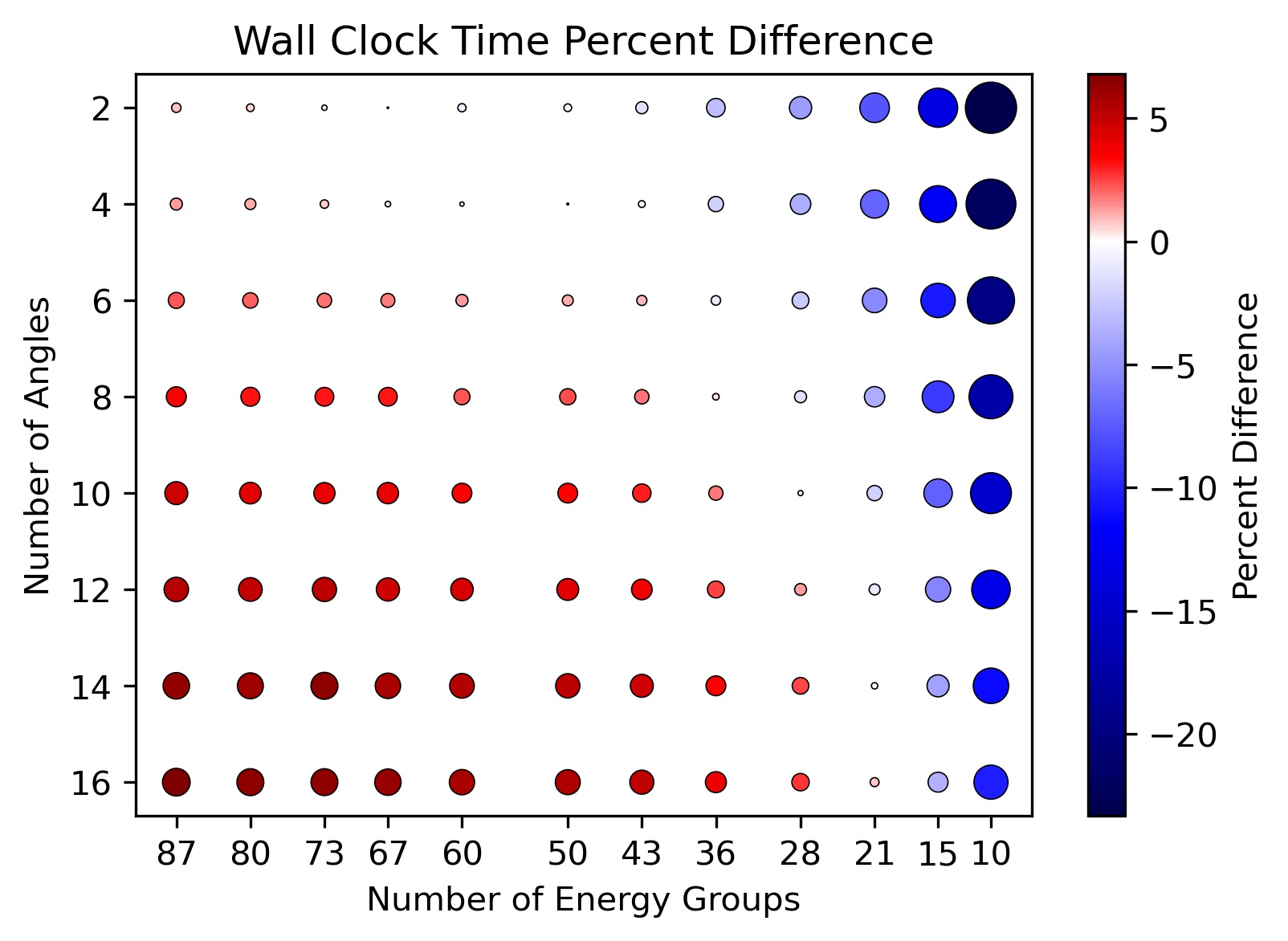} \label{fig:sphere-time-grid}} \\
    \subfloat[]{\includegraphics[width=0.52\textwidth]{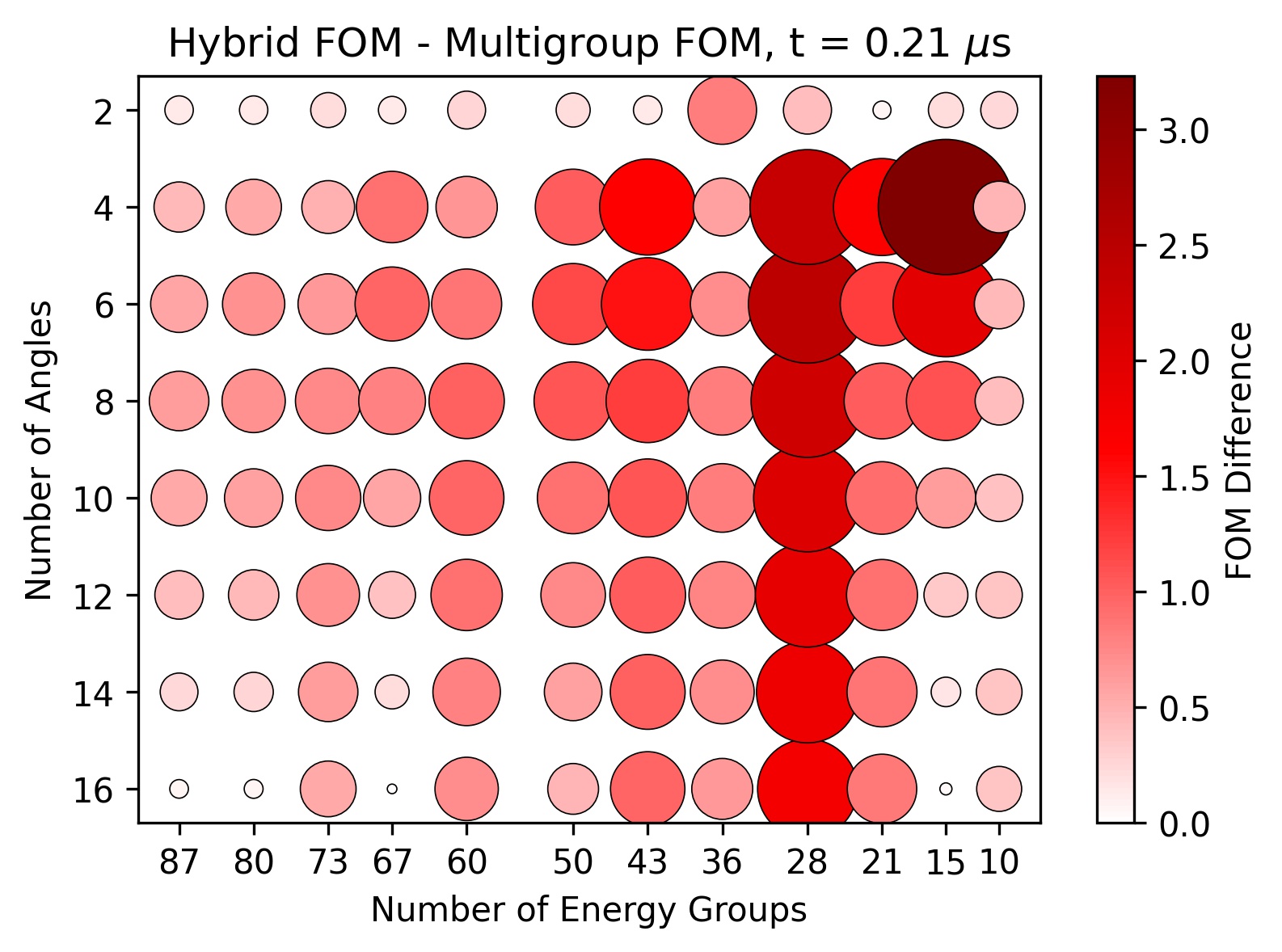} \label{fig:sphere-fom}}
    \caption{Efficiency comparison between hybrid and multigroup methods applied to the enriched and depleted uranium sphere problem across a range of discretization parameters: (a) multigroup error minus hybrid error; (b) percent difference in wall clock time; (c) hybrid FOM minus multigroup FOM. Errors in (a) are computed at $t$ = 0.21 $\mu$s with respect to a reference solution with $G = 87$ groups and $M = 32$ angles; a positive difference favors the hybrid. The errors for $\hG = 10$ are quite large and have thus been omitted to make the remaining data easier to view. For (b), the percent difference in wall clock time is $(\tau_{\rm{mg}} - \tau_{\rm{hy}}) / \tau_{\rm{mg}}$; positive values mean that the hybrid is faster. In (c), the FOM difference is computed using \eqref{eq:fom}; positive values favor the hybrid.}
    \label{fig:sphere-grid}
\end{figure}

In Fig.~\ref{fig:sphere-grid}, we compare the efficiency of the multigroup and hybrid methods in computing the fission rate density over a range of energy and angle discretization parameters.
The reference solution uses $M = 32$ angles and $G = 87$ energy groups is used for these comparisons. 
In Fig.~\ref{fig:sphere-error-grid}, we show the differences in errors of the two methods.
As in the slab problem, the hybrid performs better over a large majority of the parameters.
In Fig.~\ref{fig:sphere-time-grid}, we show the differences in wall clock time for the same set of simulations. As in the slab problem, we observe that the hybrid is generally faster for high-resolution calculations while the multigroup is faster for low-resolution calculations.
To compare the number of iterations for the full problems ($G = \hG = 87$, $M = \hM = 16$), the hybrid method uses about 6729 fewer iterations per time step than the multigroup method. 
In Fig.~\ref{fig:sphere-fom}, we plot the difference between the FOM for the hybrid method and the FOM for the multigroup method.
In all cases of the uranium sphere problem, the hybrid FOM is better. 

\begin{figure}[!ht]
    \centering
    \subfloat[]{\includegraphics[width=0.48\textwidth]{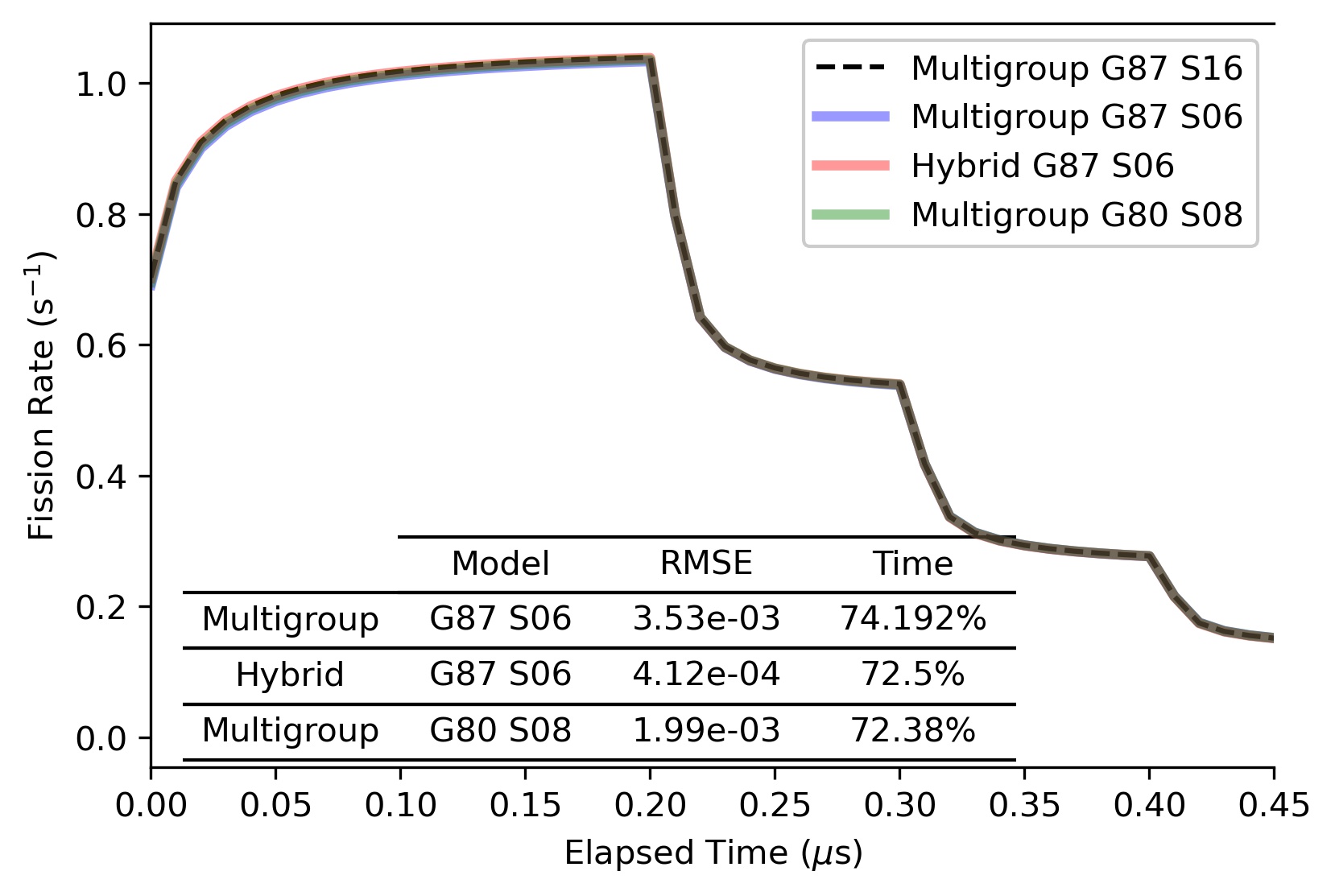} \label{fig:sphere-fission-rate-time}}
    \subfloat[]{\includegraphics[width=0.48\textwidth]{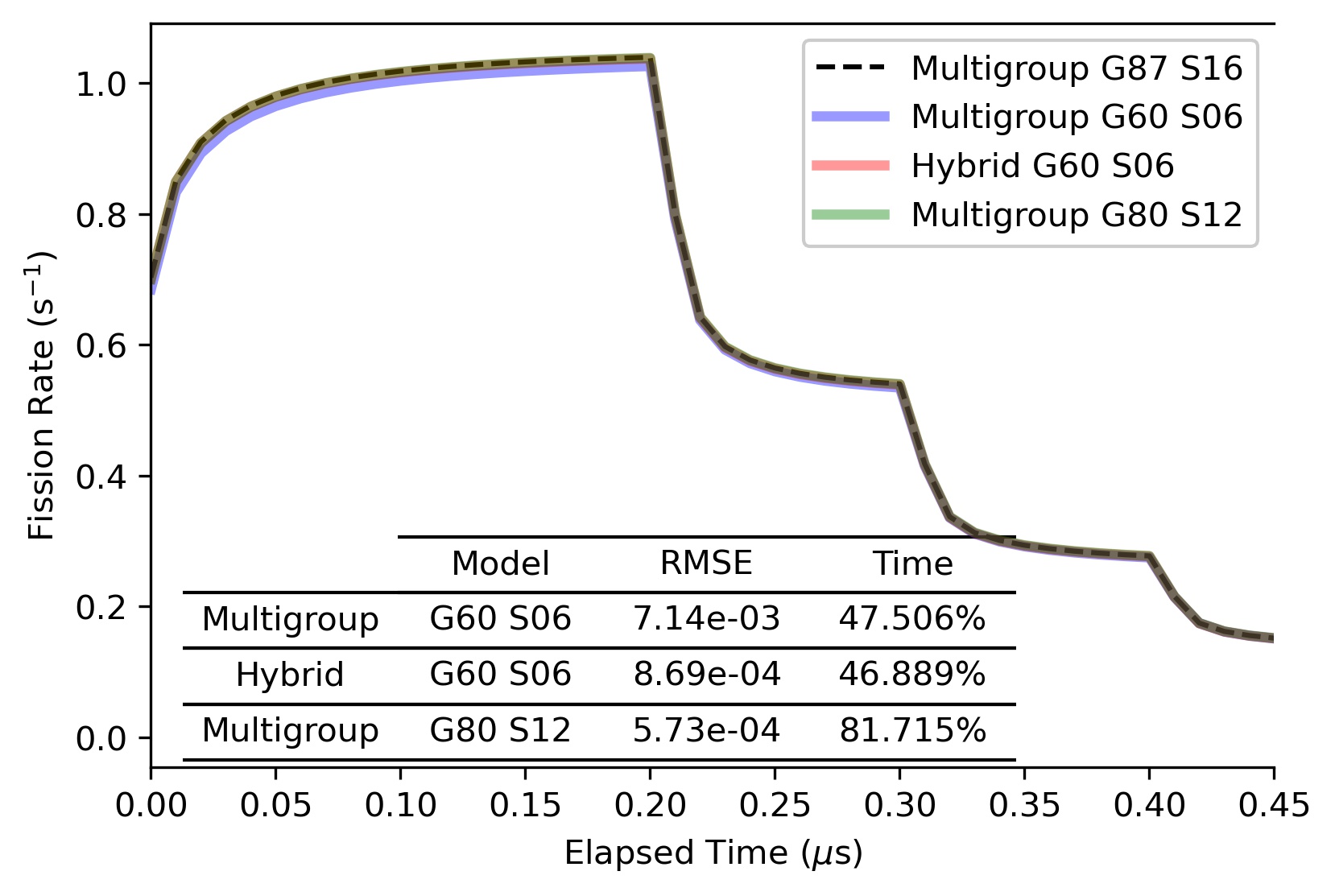} \label{fig:sphere-fission-rate-error}}
    \caption{Illustrative comparisons of the hybrid and multigroup methods in computing the fission rate for the uranium sphere problem.  For similar wall clock times, the hybrid yields smaller errors.  For similar errors, the hybrid yields smaller wall clock times.}
    \label{fig:sphere-fission-rate}
\end{figure}

In Fig.~\ref{fig:sphere-fission-rate}, we plot illustrative results from the hybrid and multigroup calculations. 
These results demonstrate that for models with similar errors, the hybrid method yields faster computational times and for models with  comparable wall clock times, the hybrid method yields more accurate results.

\subsection{Uranium oxide slab with plane source}
\begin{figure}[!ht]
    \centering
    \includegraphics[width=0.5\textwidth]{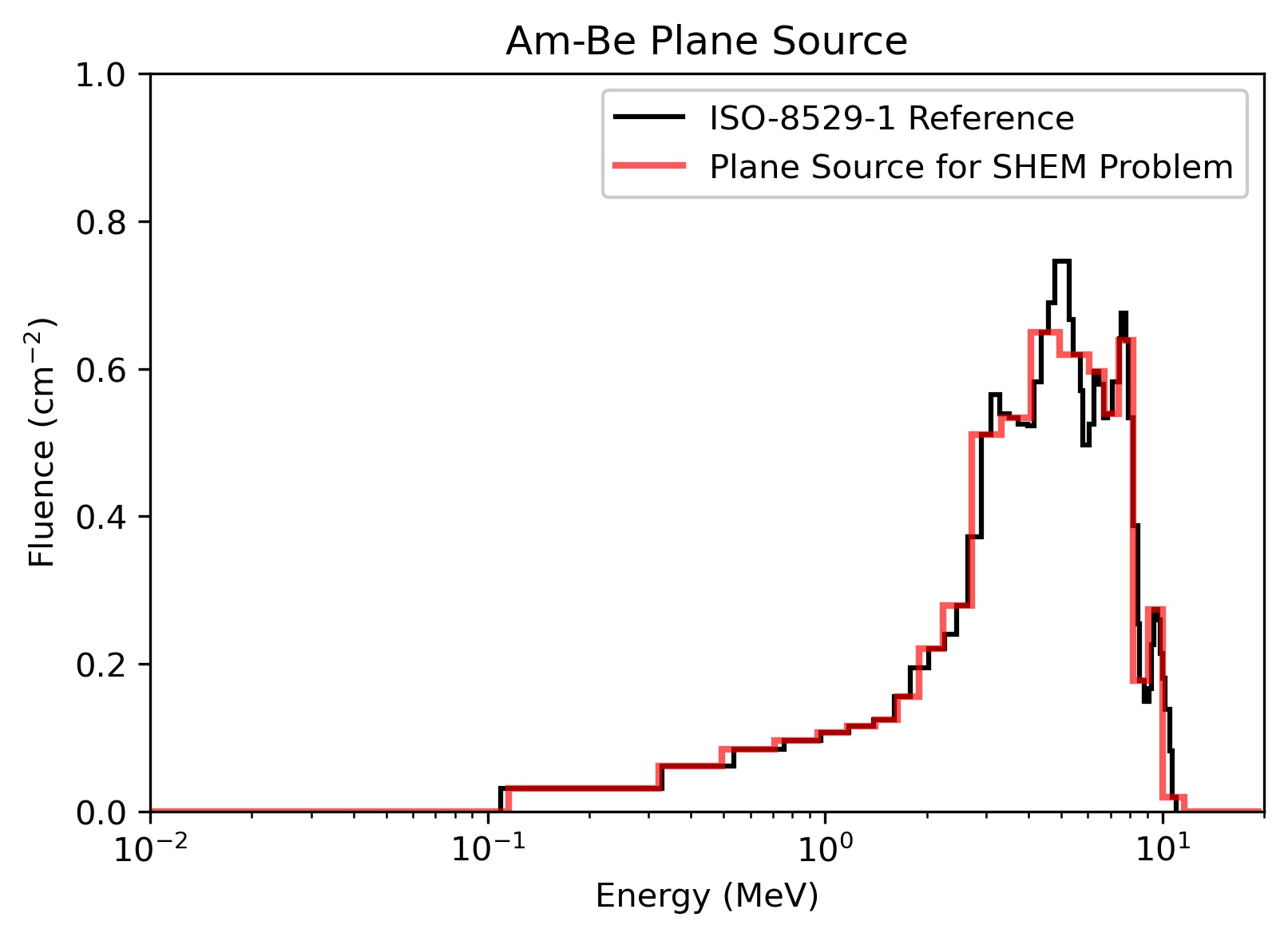}
    \caption{The plane source used in the uranium oxide Slab problem was based off the Americium-Beryllium (Am-Be) source from ISO 8529~\cite{ambe-source}. The reference source is shown by the black line while the red line is adhering the Am-Be source with the energy grid from the $G = 361$ energy group problem.}
    \label{fig:ambe-source}
\end{figure}

The last test problem involves a mixture of uranium oxide and water as it is typically found in a light water reactor.  
A one-dimensional critical width slab with $\mathbf{X} = 76.7297$ cm was used with \begin{equation} \label{eq:shem-boundary}
    b_{m,g}(\bx = 0, t) = b_{m, g}(\bx = \mathbf{X}, t) = 0 \qquand
    f_{m,g}(\bx) = 0, 
\end{equation} 
as the boundary and initial conditions.
The plane source is based off of the americium-beryllium source from the ISO 8529 standard~\cite{ambe-source} and is adjusted for the 361 energy group grid, as shown in Fig.~\ref{fig:ambe-source}.
This problem takes time step sizes of 0.01 $\mu$s with an elapsed time of 1 $\mu$s while the reference solution uses $G = 361$ energy groups and $M = 16$ angles.  The group boundaries are computed using PyNjoy~\cite{pynjoy}.  Groups are coarsened in an attempt to preserve the $k$-effective value of the slab.
The results of this more physically motivated coarsening strategy are show in Fig.~\ref{fig:energy-splitting-361}.

\begin{figure}[!ht]
    \centering
    \subfloat[]{\includegraphics[width=0.48\textwidth]{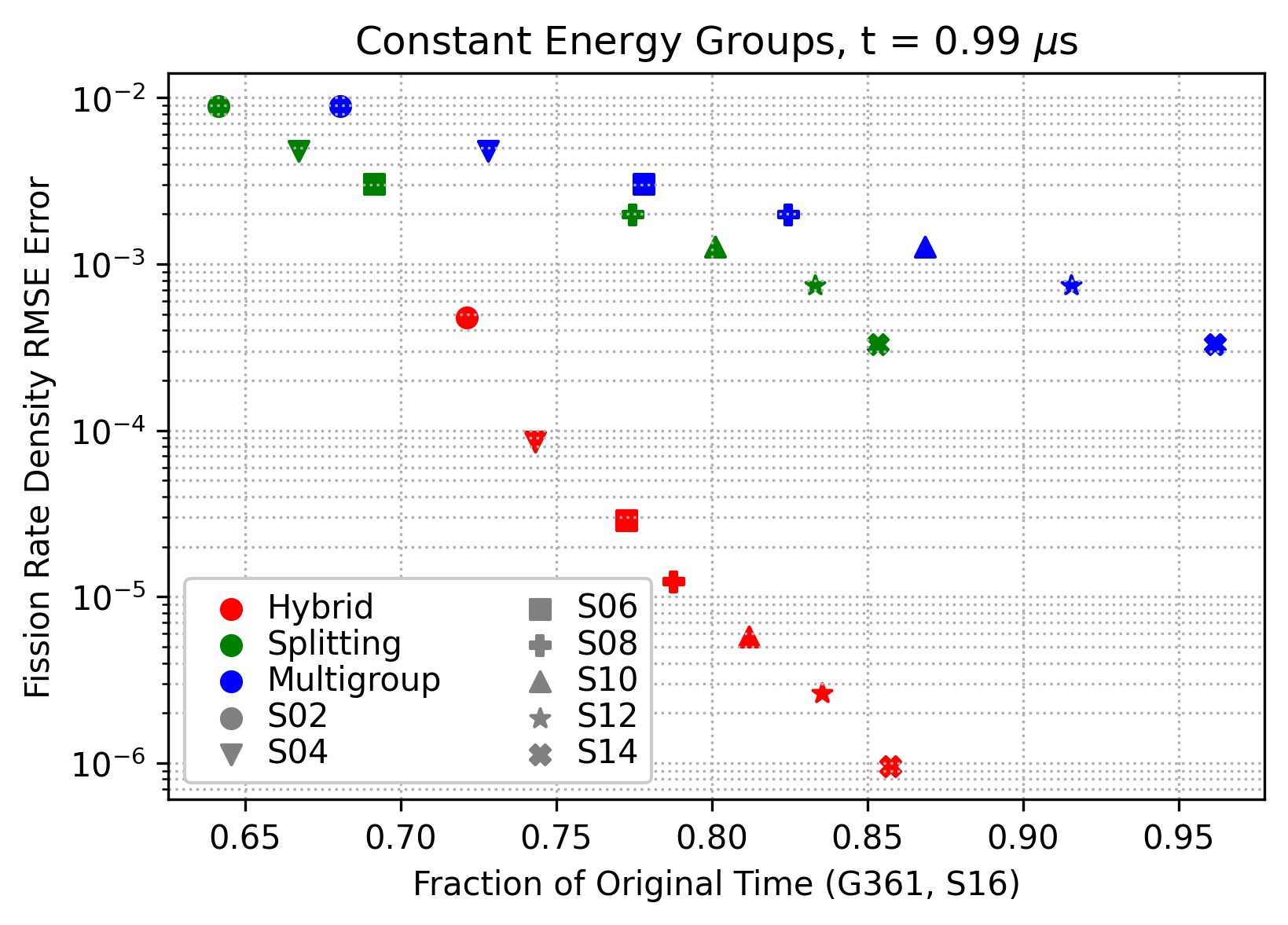} \label{fig:shem-constant-groups}}
    \subfloat[]{\includegraphics[width=0.48\textwidth]{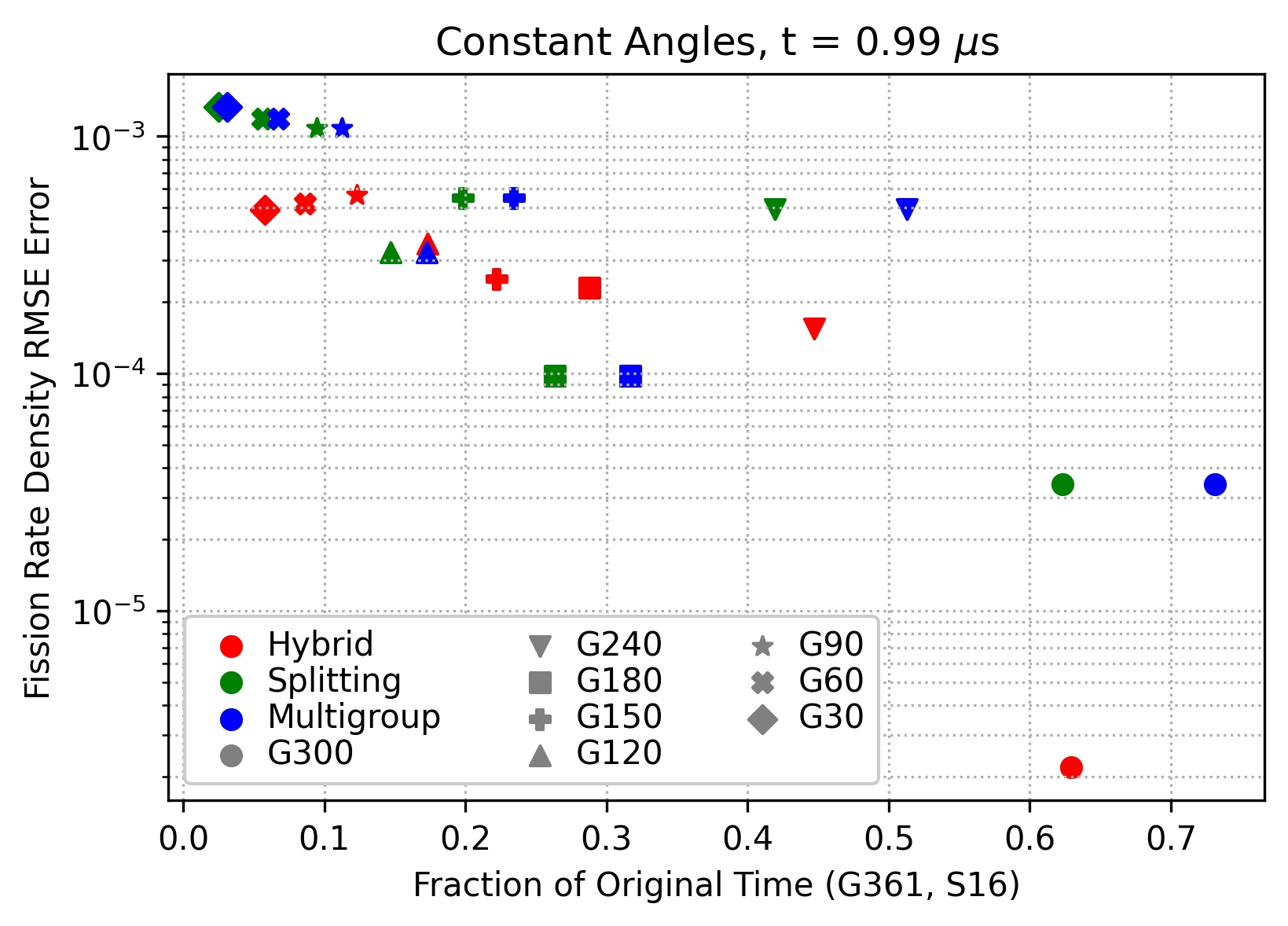} \label{fig:shem-constant-angles}}
    \caption{Comparison  the RMSE error vs. wall clock time in the fission rate density for the the uranium oxide slab at $t=0.99~\mu$s: (a) constant numbers of energy groups and varying number angles and (b) constant number of angles and varying number of energy groups. As expected, the splitting method has the same error as the multigroup method, but takes less time to compute.  The difference in wall clock time increases as $\hG$ and $\hM$ increase. The hybrid method is typically more accurate, but slightly more expensive than the splitting method. Two notable exceptions are the cases $\hG = 120$ and $\hG = 180$ in (b), which can be attributed to how the energy groups were coarsened for the multigroup method.}
    \label{fig:shem-analysis}
\end{figure}

In Fig.~\ref{fig:shem-analysis}, we compare the error versus wall clock time of the three methods described at the beginning of the section, fixing either the number of groups or the number of angles.  
The reference solution for these comparisons is a grid with $G = 361$ energy groups and $M = 16$ angles. 
In Fig.~\ref{fig:shem-constant-groups}, we vary the number of coarse angles $\hat{M} < M = 16$ while fixing $\hG = G = 361$ energy groups.  
In Fig.~\ref{fig:shem-constant-angles}, we vary the number of coarse groups $\hat{G} < G = 361$, while fixing $\hM = M = 16$ angles. 
The overall results are similar to the previous problems.  First the multigroup and splitting results yield the same errors.  However the splitting is faster and the relative difference in wall clock time increases as $\hat{M}$ or $\hat{G}$ increases.  
When compared to the splitting, the hybrid provides improved accuracy with a marginal cost increase, except when $\hG=180$ and marginally when $\hG = 120$, in which case the multigroup method performs remarkably well.

\begin{figure}[!ht]
    \centering
    \subfloat[]{\includegraphics[width=0.52\textwidth]{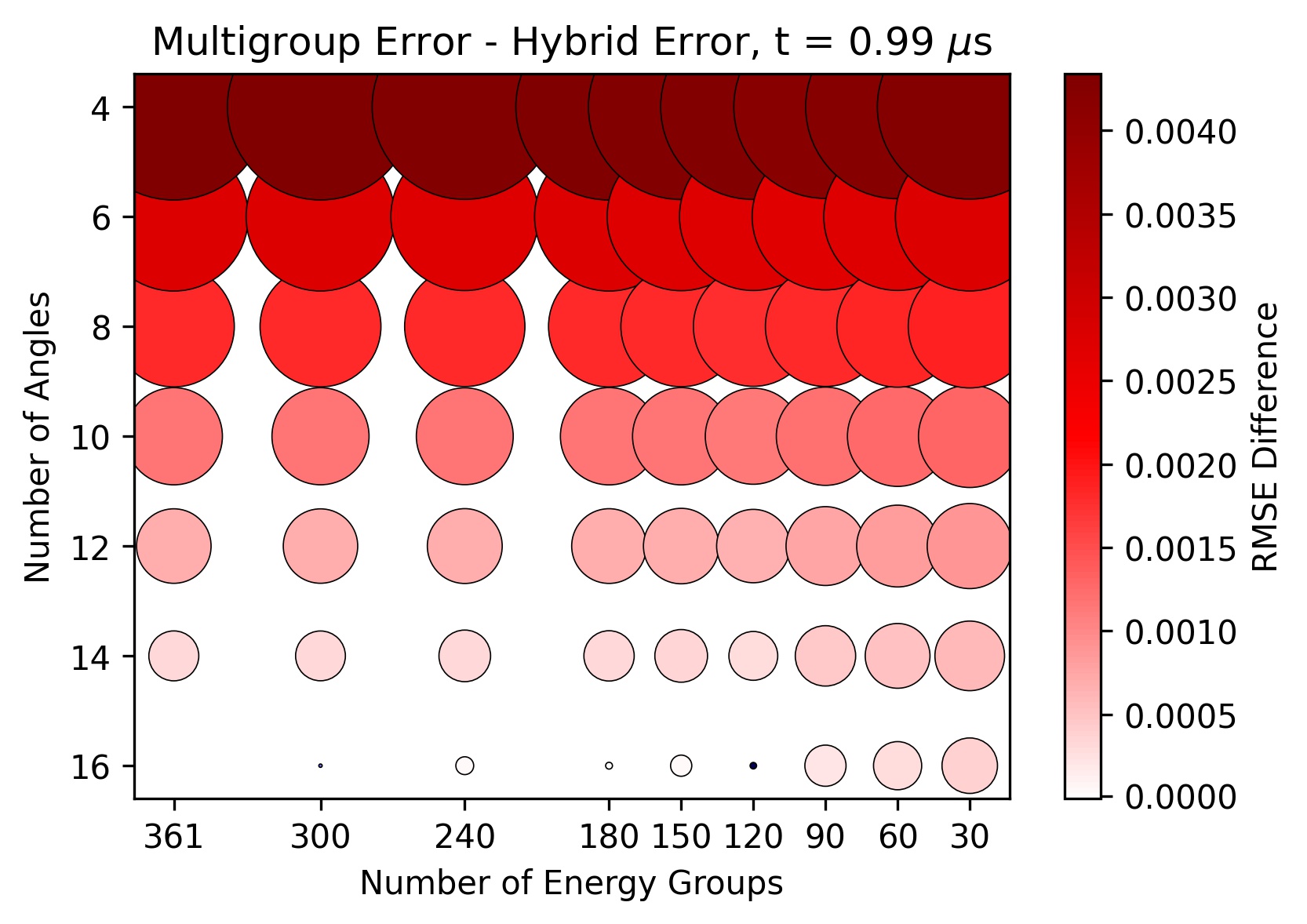} \label{fig:shem-error-grid}} 
    \subfloat[]{\includegraphics[width=0.52\textwidth]{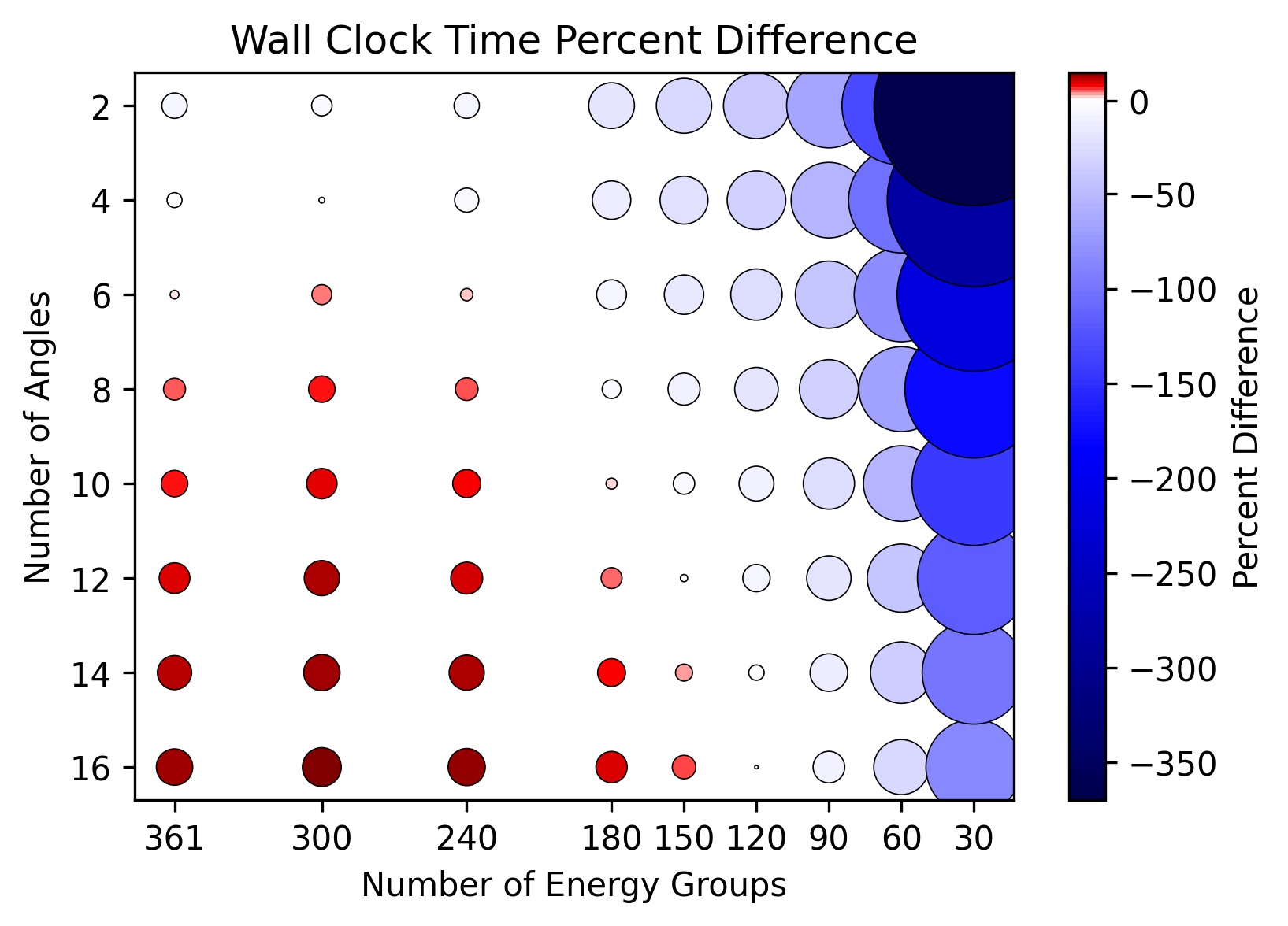} \label{fig:shem-time-grid}} \\
    \subfloat[]{\includegraphics[width=0.52\textwidth]{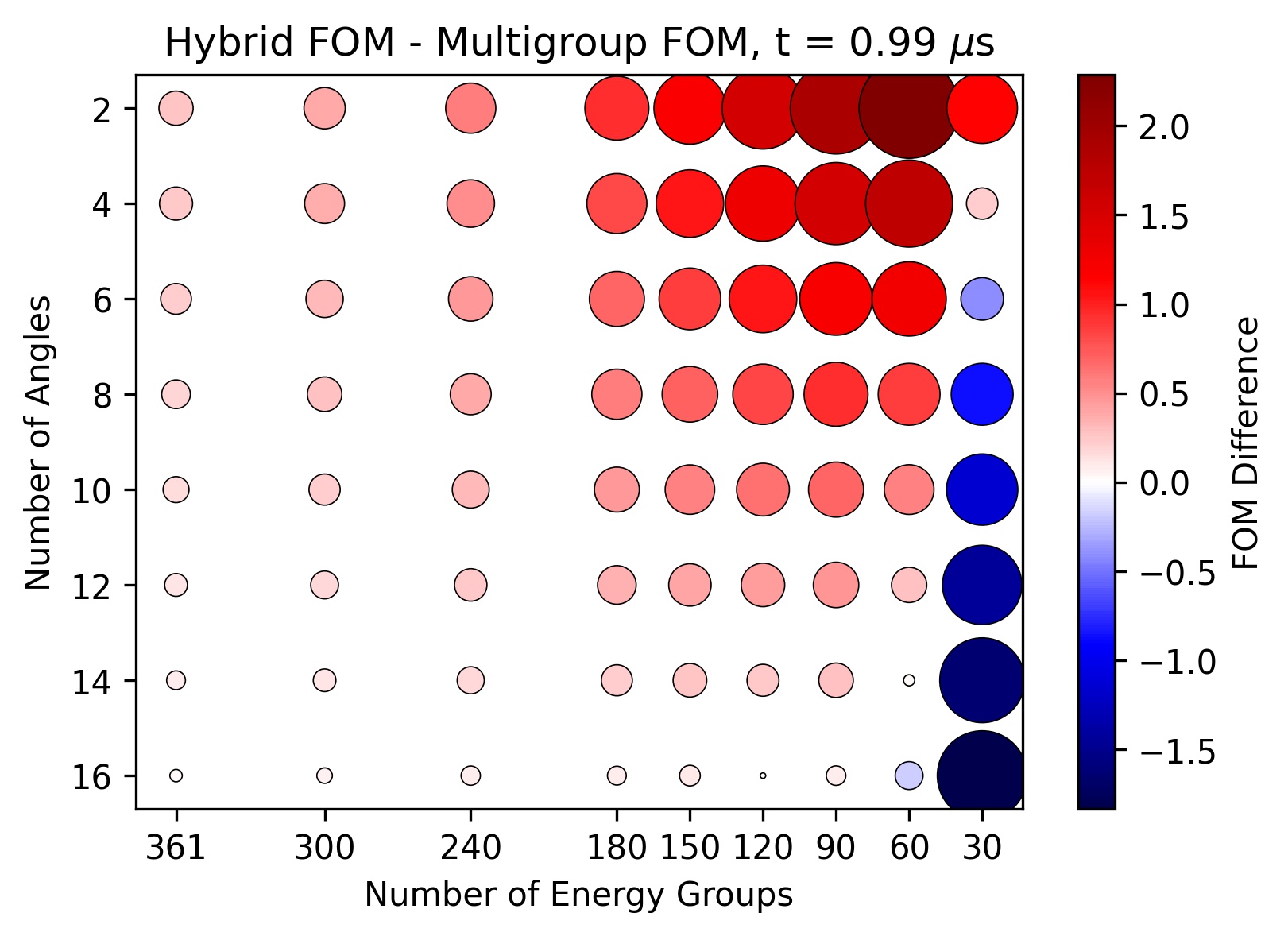} \label{fig:shem-fom}}
    \caption{Efficiency comparison between hybrid and multigroup methods applied to the uranium oxide slab problem across a range of discretization parameters: (a) multigroup error minus hybrid error; (b) percent difference in wall clock time; (c) hybrid FOM minus multigroup FOM. Errors in (a) are computed at $t$ = 0.99 $\mu$s with respect to a reference solution with $G = 361$ groups and $M = 32$ angles; a positive difference favors the hybrid.  For (b), the percent difference in wall clock time is $(\tau_{\rm{mg}} - \tau_{\rm{hy}}) / \tau_{\rm{mg}}$; positive values mean that the hybrid is faster. In (c), the FOM difference is computed using \eqref{eq:fom}; positive values favor the hybrid. Notable exceptions with cases $\hG = 120$ in (a) and $\hG = 30$ in (c) can be attributed to how the energy groups were coarsened for the multigroup method.}
    \label{fig:shem-grid}
\end{figure}

In Fig.~\ref{fig:shem-grid}, we compare the efficiency of the multigroup and hybrid methods in computing the fission rate density over a range of energy and angle discretization parameters.
The reference solution for these comparisons employs a grid with $G = 361$ energy groups and $M = 32$ angles.  In Fig.~\ref{fig:shem-error-grid}, we show the differences in errors of the two methods.
For the majority of parameters, the hybrid method performs better, especially for smaller values of $\hM$ and $\hG$. 
The error difference is not negative in the $\hG = 180$ case, as seen in Fig~\ref{fig:shem-constant-groups} because a high angular reference value is used ($M = 32$) instead of the original reference of $M = 16$.
In Fig.~\ref{fig:shem-time-grid}, we show the differences in wall clock time for the same set of simulations.  Roughly speaking, we observe that the hybrid is faster for high-resolution calculations and the multigroup is faster for low-resolution calculations. This can also be seen with the iterations needed for convergence, where the full hybrid model ($G = \hG = 361$, $M = \hM = 16$) requires about 2341 fewer iterations than the full multigroup model.
In Fig.~\ref{fig:shem-fom}, we plot the difference between the FOM for the hybrid method and the FOM for  multigroup method.
In the large majority of cases, the hybrid FOM is better, although there are some exceptions when the number of collided energy groups is low ($\hG = 30$, $60$) and the number of collided angle is high. 
In this case, the error difference at the final time step does not greatly favor the hybrid method over the multigroup method, which causes a negative FOM difference.

\begin{figure}[!ht]
    \centering
    \subfloat[]{\includegraphics[width=0.48\textwidth]{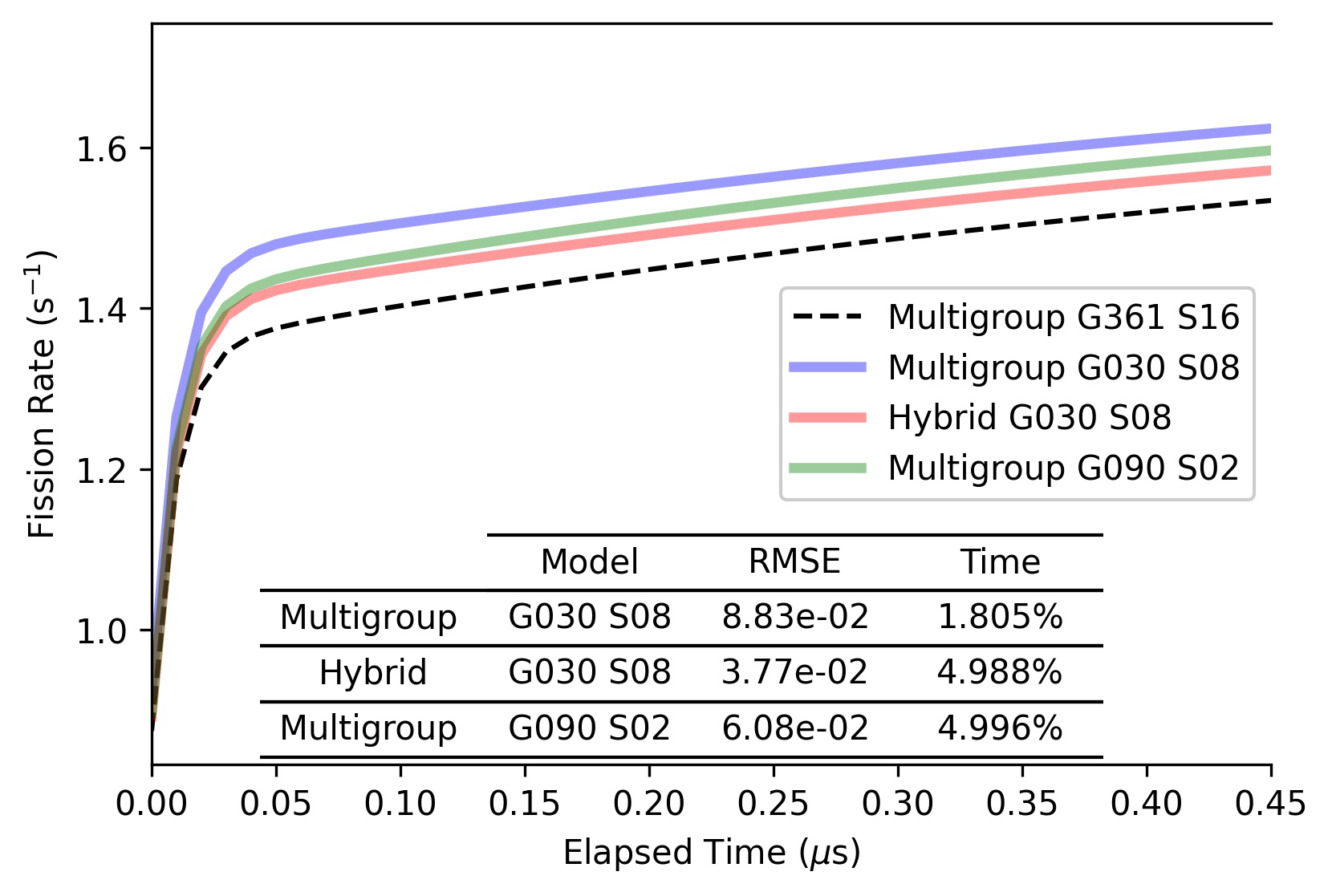} \label{fig:shem-fission-rate-time}}
    \subfloat[]{\includegraphics[width=0.48\textwidth]{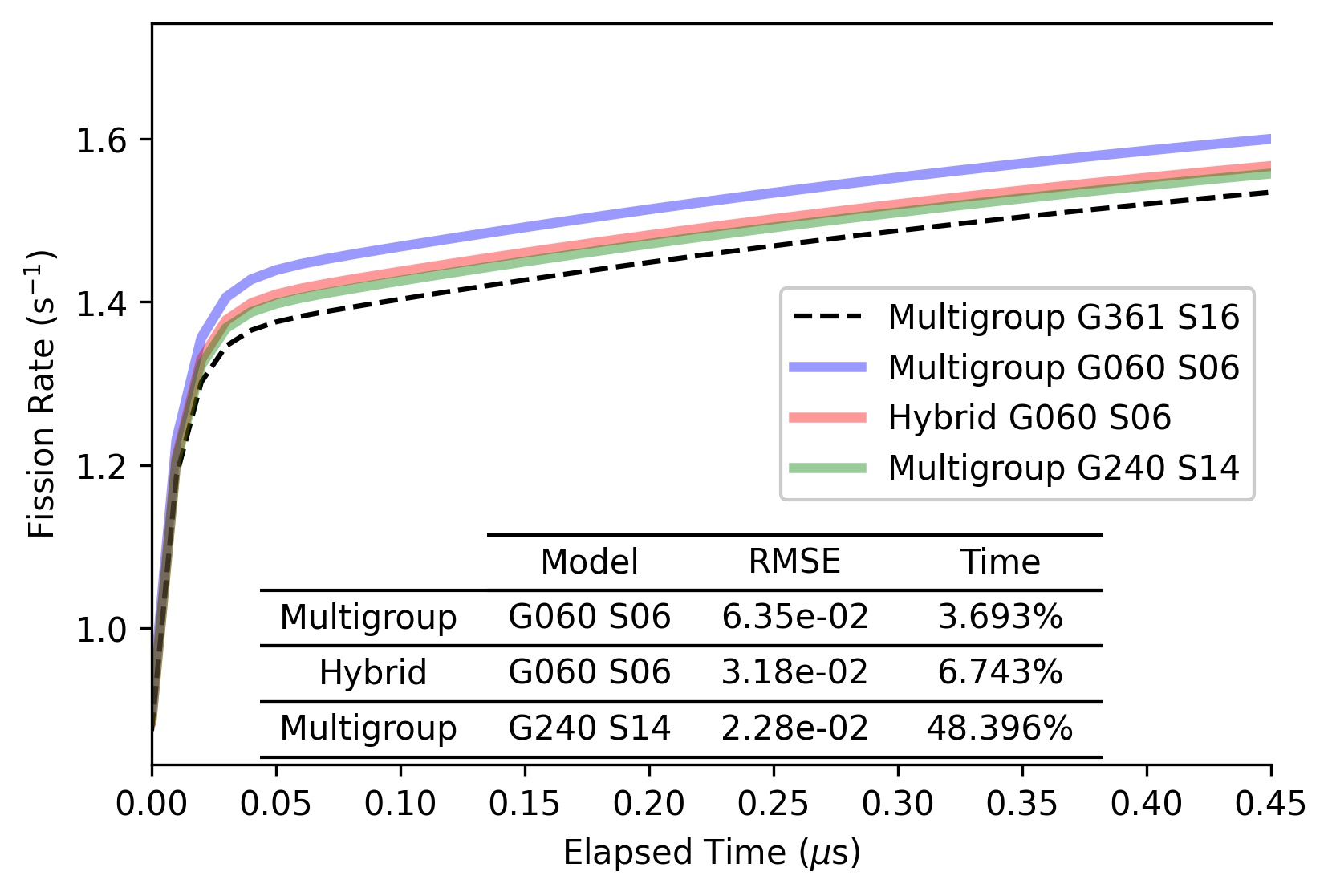} \label{fig:shem-fission-rate-error}}
    \caption{Illustrative comparisons of the hybrid and multigroup methods in computing the fission rate for the uranium oxide slab problem.  For similar wall clock times, the hybrid yields smaller errors.  For similar errors, the hybrid yields smaller wall clock times.}
    \label{fig:shem-fission-rate}
\end{figure}

In Fig.~\ref{fig:shem-fission-rate}, we plot illustrative results from the hybrid and multigroup calculations. 
As in the previous test problems, these results demonstrate that for models with similar errors, the hybrid method yields faster computational times and for models with comparable wall clock times, the hybrid method yields more accurate results.
The wall clock times are shown as a percentage of the multigroup ($G = 361$, $M = 16$) wall clock reference solution.

\section{Conclusion and future work} \label{sec:conclusion}
We have extended a discrete ordinates hybrid method of the time-dependent neutron transport equation to the multigroup setting.  The hybrid relies on a splitting of the NTE into collided and uncollided components and takes advantage of the fact that the collided component can be accurately approximating using a relatively coarse discretization in angle and energy.  The resulting discretization is typically more efficient that a single monolithic scheme.

We test the hybrid method and compare it to the usual multigroup approach that reduces computational expense coarsening the energy groups.  Except in isolated cases, the hybrid method outperforms the standard coarsening strategy.  A by-product of the hybrid approach is a faster solver, even when the hybrid and multigroup method use the same monolithic resolution. This improvement in wall clock time is attributed to the fact that the uncollided component does not require converging an inner iteration of angle; rather the inner iteration need only be performed once.  Similar results have been demonstrated in \cite{senecal2017}, where savings are realized in a systematic way by not fully converging inner iterations in a nested scheme.

There is more work to be done on hybrid approaches in the multigroup setting. Previous work on the hybrid methods for single-group problems suggests that the benefit of the hybrid with respect to angular accuracy increases in multiple spatial dimensions. We expect this behavior to carry over to the multigroup case, but this conjecture still has to be investigated. 
Previous work has also looked into the splitting the spatial dimension into collided and uncollided portions \cite{crockatt2020improvements}, but the division of the temporal dimension has not been investigated. This area should be explored, as well as the associated cost benefit of coarse collided time steps and fine uncollided time steps.
In additional, the method can be further improved by applying acceleration techniques to the collided equation and by using correction techniques to remove coarse grid error \cite{crockatt2020improvements}.

\section*{Acknowledgments}
The work of Ben Whewell and Ryan McClarren is supported by the Center for Exascale Monte-Carlo Neutron Transport (CEMeNT) a PSAAP-III project funded by the Department of Energy, grant number DE-NA003967.

The work of Cory Hauck is sponsored by the Office of Advanced Scientific Computing Research, U.S. Department of Energy, and performed at the Oak Ridge National Laboratory, which is managed by UT-Battelle, LLC under
Contract No. De-AC05-00OR22725 with the U.S. Department of Energy. The United States Government retains,
and the publisher, by accepting the article for publication, acknowledges, that the United States Government retains a non-exclusive, paid-up, irrevocable, world-wide license to publish or reproduce the published form of this
manuscript, or allow others to do so, for United States Government purposes. The Department of Energy will provide public access to these results of federally sponsored research in accordance with the DOE Public Access Plan
(http://energy.gov/downloads/doe-public-access-plan).

\bibliographystyle{styling/ans_js}
\bibliography{styling/resources}

\newpage
\appendix
\section{Algorithms for the Multigroup and Hybrid Methods} \label{sec:algorithms}

\begin{algorithm}[!ht]
	\caption{Backward Euler update of multigroup, discrete ordinates equations.}
	\label{alg:multi}
	\setstretch{1.4}
	\begin{algorithmic}[1]
	    \Require $\sig[g]{t}$, $\sig[g' \to g]{s}$, $\chi_g$, $\nu_{g'}$, $\sig[g' \to g]{f}$, $c_g$, $q_g$ \Comment{Material Properties}
	    
	    \Require $\psi_{m,g} \,^{(n-1)}$ \Comment{Solution from Previous Time Step}
	    
	    \Require $h$, $\bsOmega_m$, $w_m$ \Comment{Discretization Parameters}
	    
	    \Require $\varepsilon_G, \varepsilon_M$ \Comment{Convergence Tolerances}
	    
        \State $\Delta_G \gets 1 + \varepsilon_G$, $j \gets 0$
        
        \State $\barpsi_{g}\,^{0} \gets \displaystyle\sum_{m=1}^M w_m \psi_{m,g}\,^{(n-1)}$
        
        \While{$\Delta_G$ $> \varepsilon_G$} \Comment{Outer Iteration ($j$)}
            \For{$g = 1, \dots, G$} \Comment{Loop over Groups}
                \State $\Tilde{Q}_{g} \gets q_{g} + \displaystyle\sum_{g'=1}^{g-1} \sig[g' \rightarrow g]{s} \barpsi_{g'} \,^{j+1} +
                \displaystyle\sum_{g' = g+1}^{G} \sig[g' \rightarrow g]{s} \barpsi_{g'} \,^{j}$ 
                
                \State $\Tilde{Q}_{g} \gets \Tilde{Q}_{g} + \chi_{g} \displaystyle\sum_{g'=1}^{g-1} \nu_{g'} \sig[g' \rightarrow g]{f} \barpsi_{g'} \,^{j+1} + \chi_{g} \displaystyle\sum_{g'= g+1}^{G} \nu_{g'} \sig[g' \rightarrow g]{f} \barpsi_{g'} \,^{j}$
                
                \vspace{0.5em}
                
                \State $\Delta_M \gets 1 + \varepsilon_M$, $\quad\ell \gets 0$
                
                \State $\barpsi_{g}\,^{j+1,0} \gets \barpsi_{g}\,^{j}$

                \While{$\Delta_M > \varepsilon_M$} \Comment{Source Iteration ($\ell$)}
                    \For{$m = 1, \dots, M$} \Comment{Loop over Angles}

                        \State $\Tilde{Q}_{m,g} \gets \Tilde{Q}_{g} + \sig[g \to g]{s}\barpsi_g\,^{j+1, \ell} + \chi_{g} \nu_{g} \sig[g \to g]{f}\barpsi_g\,^{j+1, \ell} +  \dfrac{1}{c_g h} \psi_{m, g} \,^{(n-1)}$

                        \State $\psi_{m, g} \,^{j+1,\ell + 1} \gets \left(\dfrac{1}{c_g h} + \bsOmega_m \cdot \nabla + \sig[g]{t}\right)^{-1} \Tilde{Q}_{m, g}$ \Comment{Transport Sweep}
                        
                    \EndFor
                    
                    \State $\barpsi_{g}\,^{j+1, \ell+1} \gets \displaystyle\sum_{m=1}^M w_m \psi_{m,g}\,^{j+1, \ell+1}$
                    
                    \State $\Delta_M$ $\gets \displaystyle \left| \left| \dfrac{\barpsi_{g} \,^{j+1, \ell + 1} - \barpsi_{g} \,^{j+1, \ell}}{\barpsi_{g} \,^{j+1, \ell + 1}} \right| \right|_{2}$
                    
                    \State $\ell \gets \ell + 1$
                \EndWhile
                \State $\psi_{m,g}\,^{j+1} \gets \psi_{m,g}\,^{j+1,\ell}\qquad\barpsi_{g}\,^{j+1} \gets \barpsi_{g}\,^{j+1, \ell}$
            \EndFor
            \State $\Delta_G$ $\gets \displaystyle \left| \left| \dfrac{\barpsi \,^{j + 1} - \barpsi \,^{j}}{\barpsi \,^{j + 1}} \right| \right|_{2}$ \Comment{$\barpsi \,^{j+1} = \left[ \, \barpsi_1 \,^{j+1}, \dots, \barpsi_G \,^{j+1} \, \right]$}
            \State $j \gets j + 1$
        \EndWhile 
        \State $\psi_{m,g}\,^{(n)} \gets \psi_{m,g}\,^{j},\qquad
        \barpsi_{g}\,^{(n)} \gets \barpsi_{g}\,^{j}$
        \State \Return $\psi_{m,g} \,^{(n)}$, $\barpsi_g \,^{(n)}$
	\end{algorithmic} 
\end{algorithm}

\newpage

\begin{algorithm}[!ht]
	\caption{Backward Euler update of the collision-based hybrid method for multigroup discrete ordinates equations.}
	\label{alg:hybrid}
	\setstretch{1.4}
	\begin{algorithmic}[1]
	
    	\Require $\sig[g]{t}$, $\sig[g' \to g]{s}$, $\chi_{g}$, $\nu_{g'}$, $\sig[g' \to g]{f}$, $c_g$, $q_g$, \Comment{Uncollided Material Properties}
    	
    	\Require $\hsig[\hg]{t}$, $\hsig[\hg' \to \hg]{s}$, $\hat \chi_{\hg}$, $\hat \nu_{\hg'}$, $\hsig[\hg' \to \hg]{f}$, $c_{\hg}$ \Comment{Collided Material Properties}
    	
    	\Require $\psi^{(n-1)}$ \Comment{Solution from Previous Time Step}
    	
    	\Require $h$, $\bsOmega_m$, $w_{m}$, $\hat{\bsOmega}_{\hm}$, $\hw_{\hm}$, $\Delta E_g$, $\Delta \hE_{\hg}$ \Comment{Discretization Parameters}
    	
    	\Require $\varepsilon_G$, $\varepsilon_M$ \Comment{Convergence Tolerances}
    	
	    \State $\psiu_{m,g} \,^{(n-1)} \gets \psi_{m,g} \,^{(n-1)} \qquad \qu_{g} \gets q_{g}$
        \For{$g = 1, \dots, G$} \Comment{Uncollided Flux Update}
            \For{$m = 1, \dots, M$} 
                \State $\psiu_{m,g} \gets \left(\dfrac{1}{c_g h} + \bsOmega_m \cdot \nabla + \sig[g]{t}\right)^{-1} \left(\qu_{g} + \dfrac{1}{c_g h} \psiu_{m, g} \,^{(n-1)} \right)$ \Comment{Transport Sweep}
            \EndFor
        \EndFor
        \State $\barpsiu_g \gets \displaystyle\sum_{m = 1}^{M} w_{m} \psiu_{m,g}$
        
        \vspace{0.5em}
        \State $\qc_{\hg} \gets$
        $\displaystyle\sum_{g = \alpha_{\hg}+1}^{\alpha_{\hg+1}} \displaystyle\sum_{g'=1}^{G} \sig[g' \rightarrow g]{s} \barpsiu_{g'} + \displaystyle\sum_{g=\alpha_{\hg}+1}^{\alpha_{\hg+1}} \chi_{g} \displaystyle\sum_{g'=1}^{G} \nu_{g'} \sig[g' \rightarrow g]{f} \barpsiu_{g'}$
        \Comment{Collided Source}
        \vspace{0.5em}
        
        \State $\barpsic_{\hm, \hg} \gets$ Algorithm~\ref{alg:multi} with: \Comment{Collided Flux Update} 
        
        $\sig[g]{t} \gets \hsig[\hg]{t}$, $\sig[g' \to g]{s} \gets \hsig[\hg' \to \hg]{s}$, $\chi_{g} \gets \hat \chi_{\hg}$, $\nu_{g'} \gets \hat \nu_{\hg'}$, $\sig[g' \to g]{f} \gets \hsig[\hg' \to \hg]{f}$, $c_g \gets c_{\hg}$
        
        $q_g \gets \qc_{\hg}$, $\bsOmega_m \gets \hat \bsOmega_{\hm}$, $w_m \gets \hw_{\hm}$, $\psi^{(n-1)} \gets 0$, $\varepsilon_G \gets \varepsilon_G$, and $\varepsilon_M \gets \varepsilon_M$
        
        \vspace{0.5em}
        \State $\qt_{m,g} \gets \dfrac{\Delta E_g}{\Delta \hE_{\hg}} \left[\qc_{\hg} +
        \displaystyle\sum_{\hg'=1}^{\hG} \sig[\hg' \rightarrow \hg]{s} \barpsic_{\hg'} + \chi_{\hg} \displaystyle\sum_{\hg'=1}^{\hG} \nu_{\hg'} \sig[\hg' \rightarrow \hg]{f} \barpsic_{\hg'} \right]$
        \Comment{Total Source}
        
        \vspace{0.5em}
        \For{$g = 1, \dots, G$} \Comment{Total Flux Update}
            \For{$m = 1, \dots, M$}
                \State $\psit_{m, g} \gets \left(\dfrac{1}{c_g h} + \bsOmega_m \cdot \nabla + \sig[g]{t}\right)^{-1} \qt_{m,g}$ \Comment{Transport Sweep}
            \EndFor
        \EndFor
        \State $\psi_{m, g} \,^{(n)} \gets \psit_{m, g} \qquad \barpsi_g \,^{(n)} \gets \displaystyle\sum_{m = 1}^{M} w_{m} \psit_{m,g}$
        \State \Return $\psi_{m,g} \,^{(n)}$, $\barpsi_{g} \,^{(n)}$
	\end{algorithmic} 
\end{algorithm}

\end{document}